\documentclass[10pt,pre,aps,twocolumn,floatfix,nofootinbib]{revtex4-1}
\usepackage{graphicx}
\usepackage{amssymb}
\usepackage{amsmath}

\begin{document}

\title{Using operator covariance to disentangle scaling dimensions in lattice models}

\author{Anders W.~Sandvik}
\email{sandvik@bu.edu}
\affiliation{Department of Physics, Boston University, 590 Commonwealth Avenue, Boston, Massachusetts 02215}
\affiliation{School of Physical and Mathematical Sciences, Nanyang Technological University, Singapore}

\begin{abstract}
In critical lattice models, distance ($r$) dependent correlation functions contain power laws $r^{-2\Delta}$ governed by scaling 
dimensions $\Delta$ of an underlying continuum field theory. In Monte Carlo simulations and other numerical approaches, the leading 
dimensions can be extracted by data fitting, which can be difficult when two or more powers contribute significantly. Here a method 
utilizing covariance between multiple lattice operators is developed where the $r$ dependent eigenvalues of the covariance matrix
reflect scaling dimensions of individual field operators. This disentangling of scaling dimensions is demonstrated explicitly for
conformal field theories. The computational scheme is first tested on the critical point of the two-dimensional Ising model,
where the two primary scaling dimensions and their respective two lowest descendant dimensions are extracted. The three-dimensional
Ising model is studied next, revealing the two relevant primaries and their lowest descendants to high precision. For a more challenging
case, the tricritical Ising point in two dimensions is studied with the Blume-Capel (diluted Ising) model. Here the scaling dimensions of
all three fully symmetric (under lattice point group and spin-inversion transformations) primary operators are successfully isolated along
with the leading descendants. The eigenvectors in the space of the two relevant primary operators are also studied and give useful information
on the boundary between the ordered and disordered phases in the neighborhood of the tricritical point. Finally, the crossover from regular
to tricritical Ising scaling is investigated on several points on the phase boundary of the Blume-Capel model away from its tricritical
point. The scaling of the eigenvalues corresponding to tricritical descendant operators are found to be remarkably stable even far from the
tricritical point. The covariance method represents a simple extension of standard analysis of correlation functions and can significantly
enhance the utility of Monte Carlo simulations and other computational methods in studies of classical and quantum criticality.
\end{abstract}

\date{\today}

\maketitle

\section{Introduction}

The most direct bridge between a critical point in a lattice model and its continuum field theoretical description is through the spectrum
of scaling dimensions $\Delta_i$. In numerical studies, e.g., Monte Carlo (MC) simulations, the dimensions are accessible in correlation 
functions, which contain contributions decaying with distance $r$ as $r^{-2\Delta_i}$ for each field operator $i$ compatible with the symmetries 
of the lattice operator used. For a system in $d$ dimensions (space-time dimensions of a quantum system) the scaling dimensions of relevant
operators ($\Delta_i<d$) in different symmetry sectors are related to the conventional critical exponents. While these relevant scaling
dimensions are typically of primary interest, irrelevant operators ($\Delta_i>d$) can produce significant corrections that complicate the
analysis of correlation functions. To more completely characterize a critical point, it is also useful in its own right to determine some of
the irrelevant scaling dimensions. Modern conformal field theory (CFT) techniques, in particular the numerical bootstrap \cite{poland19} and
fuzzy sphere approaches \cite{zhu23}, can access an extended range of scaling dimensions. It can be desirable to extract both relevant and
irrelevant dimensions also in lattice simulations, e.g., in order to determine whether or not the low-energy properties of a system are
described by a CFT.

Indeed, some of the most challenging problems in contemporary condensed matter physics involve the problem of matching numerical simulation results to CFTs
or other quantum field theories through scaling dimensions, often with the added complexity of multicritical behavior where correlation functions contain
contributions from more than one relevant field. The case of the putative $d=2+1$ deconfined quantum critical point \cite{senthil04} is
perhaps the most prominent unresolved example. Here the precise matching of multiple power law decays of lattice correlators
\cite{kaul12,block13,nahum15,sreejith19,zhao20,takahashi24} to emerging CFT data \cite{nakayama16,li18,zhou24,zou25,chester24,chen24a,chen24b,li26}
is crucial for building a consensus on the ultimate nature of the quantum phase transition. A more mundane but also very important class of
problems with two relevant scaling dimensions is that of liquid-gas phase transitions. These are generically of Ising type \cite{kadanoff67} but,
because the $Z_2$ symmetry is only emergent, require tuning of both the pressure and the temperature to reach the critical point (unlike the Ising
model, in which the exact $Z_2$ symmetry implies that the critical point is located at zero field and only the temperature needs to be tuned).
The presence of two relevant scaling fields makes it challenging to extract the critical point and the phase boundary in its neighborhood
\cite{yarmo17,yarmo18}.

\begin{figure}[t]
\includegraphics[width=84mm]{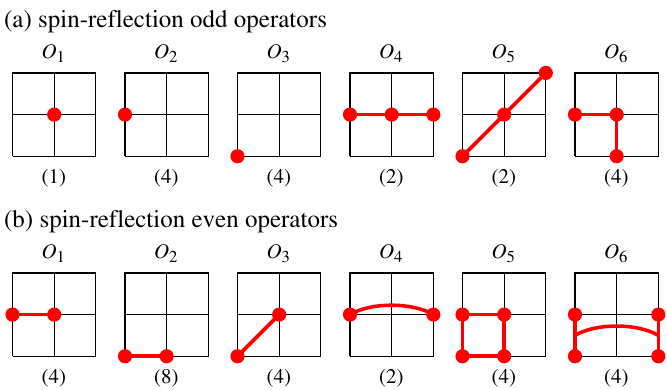}
\caption{Ising spin operators defined on cells of $3\times 3$ spins, with (a) and (b) corresponding to $Z_2$ antisymmetry and symmetry, 
respectively. The red dots depict Ising operators $\sigma_i$, with lines connecting two or more operators implying products. Point group
symmetrization is further carried out by adding all different operators resulting from rotations about the cell center and reflection
about the central horizontal or vertical axis. The total number $n_i$ of such terms is indicated beneath each of the cells.}
\label{ops_is}
\end{figure}

In this article a simple method for disentangling of different power-law contributions to numerically computed correlation functions is proposed.
The focus is on lattice models, but generalization to continuum systems will be straight-forward. Defining a set of a small number of lattice operators
$O_i({\bf x})$, $i=1,\ldots,N$ (with $N < 10$ typically) on cells centered on position  ${\bf x}$,
the full covariance matrix of these operators 
\begin{equation}
C_{ij}({\bf r}) = \langle O_i({\bf x}_1) O_j({\bf x}_2)\rangle - \langle O_i\rangle\langle O_j\rangle,~~~{\bf r} ={\bf x}_2 - {\bf x}_1,
\end{equation}
is evaluated versus the separation ${\bf r}$. In MC simulations, the expectation values include averaging over equivalent pairs of lattice
sites $({\bf x}_1,{\bf x}_2)$ and $\langle O_i\rangle$ is also a spatial average. By diagonalizing the symmetric covariance matrix, a 
set of eigenvalues $D_i({\bf r})$ is obtained. It will be shown that, for sufficiently large separation $r$ in a system described by a CFT
(or other critical low-energy theory), these eigenvalues decay with different powers of $r$ that correspond to the scaling dimensions of primary and
descendant operators compatible with the symmetries of the operators used. Examples of lattice operators defined on cells of $3\times 3$ Ising spins
are shown in Fig.~\ref{ops_is} and will be used to test the covariance method with the standard two-dimensional (2D) classical Ising model.
These operators are all fully symmetric under the relevant point group transformations of the cells, thus contain only CFT operators with
Lorentz spin $l=0$. The operators in Figs.~\ref{ops_is}(a) and \ref{ops_is}(b) are antisymmetric and symmetric, respectively, under
spin inversion ($Z_2$ symmetry), as follows from the number of operators, even or odd, in each product.
Using covariance matrices involving these operators, MC results for the 2D Ising model can reproduce the known scaling dimensions of the primary operators
as well as their first two descendants. Likewise, for the 3D Ising model, using operators similar to those in Fig.~\ref{ops_is}, the two relevant primaries
and their leading descendants are well resolved. The largest of the reproduced scaling dimensions corresponds to a power-law decay as fast as $r^{-10}$,
which is far beyond practical resolution with standard analysis of individual correlation functions.

The covariance method is particularly useful for analyzing multicritical points, where two or more primary operators are relevant. To test
such a case, the 2D Blume-Capel (BC) model will be studied at its tricritical Ising point. The BC model, or diluted Ising model, has a
three-state degree of freedom $\sigma_i \in \{-1,0,+1\}$ on each site $i$ and the Hamiltonian is
\begin{equation}
H = - \sum_{\langle ij\rangle}\sigma_i\sigma_j + \lambda \sum_i \sigma_i^2 ~\equiv~ E + \lambda F,
\label{bcham}
\end{equation}
here on the square lattice. For positive $\lambda \alt 1.966$, the model undergoes a continuous Ising transition into a ferromagnetic state,
with the critical temperature $T_c$ decreasing as $\lambda$ is increased, while for larger $\lambda \le 2$ the transition is first-order. The
location of the tricritical point $(\lambda_{\rm t},T_{\rm t})$ separating these behaviors has been estimated in several works
\cite{landau81,landau86,beale86,wilding96,deng05,silva06,plascak13,kwak15}, with the most precise reported parameters being
$\lambda_{\rm t} \approx 1.965815$, $T_{\rm t} \approx 0.608578$ from transfer-matrix calculations \cite{deng05}. This point belongs 
to the tricritical Ising universality class, having two relevant fully symmetric perturbations (operators) whose scaling dimensions are
$\Delta_0=1/5$ and $\Delta_1=6/5$ \cite{nienhuis82,yellowbook}. Reproducing these values will be important but, as it turns out, a third
irrelevant primary and some of the irrelevant dimensions can also be identified. Moreover, the method produces the linear combinations of the
two operators $E$ and $F$, defined in Eq.~(\ref{bcham}), corresponding to the relevant CFT operators. One of these eigenvectors gives the tangent
of the phase boundary at the tricritical point, which here is determined to better precision than in previous works. This ability to
reliably compute the eigenvectors in the space of two tuning parameters should be of great utility also in studies of liquid-gas phase transitions.

Operator covariance is often used in particle physics to extract masses \cite{lucher90,peardon97,peardon99,alpha09,liu12}. This method is 
in part similar to the scheme developed here, but there are important differences. In particle physics, exponential decays in imaginary (Euclidean)
time are used to extract the energy levels of a discrete spectrum, as opposed to the power law decays extracted here that correspond to a gapless
spectrum in the limit when the system length $L \to \infty$. In other words, in particle physics covariance matrices are computed versus imaginary
time separation $\tau$ in a system where the time dimension $L_\tau =1/T$, $T$ being the temperature, is much larger than the spatial length
$L$. The ability of the eigenvalues to correctly separate the different exponential decays corresponding to different masses (discrete energy
levels) was proven in that limit \cite{lucher90}, and many aspects of the method are more involved than the scheme presented here; in
particular the choice of suitable operators and the use of a reference time in a generalized eigenvalue problem.

In the present case, all dimensions in a classical system can be set equal, and in a quantum system $L_\tau$ can be of order $L$. In a critical
quantum system with an underlying CFT description and $L_\tau \gg L$ imposed, the different power-law decays corresponding to CFT
scaling dimensions would be applicable at time $\tau$ only up to $\tau \propto L$, not at $\tau \gg L$, where instead exponential decays
corresponding to the finite-size discreteness of the spectrum would apply. Thus, the present method extends the application of covariance
eigenvalues to the different domain of classical and quantum criticality. Analytical arguments
will be presented for the proper disentangling of the scaling dimensions. The numerical examples will cover only classical systems, but
the application to quantum systems is straight-forward and will be considered elsewhere \cite{lin26}.

The outline of the rest of the article is as follows. In Sec.~\ref{sec:cft}, disentangling of scaling dimensions by the covariance method is first
discussed in the framework of ideal lattice operators (in general impossible to contruct explicitly) realizing a single CFT operator. It is argued that
unique linear combinations $\tilde Q_{pi}$ of these operators $Q_{pi}$ belonging to a given symmetry sector $p$ can always be found that are orthogonal
in the correlation sense, i.e., $\langle \tilde Q_{pi}({\bf r})\tilde Q_{pj}({\bf 0})\rangle=0$ for $i \not= j$, and further that the decay with $r$ of
$i=j$ correlators is governed purely by $\Delta_i$; $\langle \tilde Q_{pi}({\bf r})\tilde Q_{pi}({\bf 0})\rangle=r^{-2\Delta_i}$. Given that simple lattice
operators (of the kind that can be easily constructed in practice) within a given symmetry sector can be written as linear combinations of the ideal
operators $\tilde Q_{pi}$ (including all $p$ compatible with the symmetries), it is then also true that
diagonalization of an $N\times N$ covariance matrix, Eq.~(\ref{cij}), produces eigenvalues whose asymptotic decays are controlled by the $N$ smallest
scaling dimensions. This assertion is exemplified by results for the 2D classical Ising model in Sec.~\ref{sec:is} and the 3D Ising model in Sec.~\ref{sec:is3},
where analogies with quantum systems in 2+1 dimensions are also made. Tricriticality in the 2D BC model is studied in Sec.~\ref{sec:bc}. In
Sec.~\ref{crossover} the crossover from tricritical to conventional Ising scaling dimensions in the neighborhood of the tricritical point
is investigated. The results and conclusions are further discussed in Sec.~\ref{sec:discuss}.

\section{Effective Correlation orthogonality} \label{sec:cft}

The ability of the covariance method to deliver scaling dimensions for a low-energy field theory from correlation functions computed with a lattice
Hamiltonian (or some other regularized model) will for definiteness here be framed within CFT terminology, though systems described by other field theories
can be similarly treated. A CFT is characterized by a set of primary operators $Q_{p0}({\bf r})$ and their descendants $Q_{pi}({\bf r})$ obtained from
the $i$:th derivatives of $Q_{p0}({\bf r})$. For a $d$-dimensional quantum system defined in the appropriate conformal geometry, for each conformal
multiplet $p$ the state-operator correspondence implies a tower of equally spaced eigenstates $\epsilon_i$ of a corresponding Hamiltonian $H$ defined only
on the system in the $d-1$ spatial dimensions. The energy gaps $\epsilon_i-\epsilon_0$ are proportional to the spectrum of scaling dimensions $\Delta_i$
of the CFT, with the constant of proportionality being of order the inverse system size; $\epsilon_i -\epsilon_0 \propto \Delta_i/L$.
The conformal geometry for $d=1+1$ is an infinite cylinder \cite{cardy82}, which for a quantum system can be realized with $H$ defined on a ring of
circumference $L$ at temperature $T \to 0$. The imaginary-time evolution operator ${\rm e}^{-\tau H}$ for $\tau \in [0,\beta]$, with $\beta=T^{-1}$,
defines the second dimension, which is normally also taken to be periodic (as in quantum statistical mechanics).

The infinite cylinder (or torus with $\beta \gg L$) geometry is easy to implement in lattice calculations for systems (Hamiltonians) with potential
CFT descriptions in one space dimension, and useful scaling dimensions can be extracted from low-lying levels $\epsilon_i$ \cite{feiguin07,suwa15,wang22}.
Though a critical point of a given microscopic model will only have a precise CFT description in the infinite-size limit, the flow of the energy levels
$\epsilon_i$ (or the scaling dimensions $\Delta_i$ extracted from some other quantity) versus the system size is normally well behaved (being controlled by
subleading corrections from irrelevant perturbations) and can be extrapolated. However, in $d=2+1$ dimensions, the periodic boundary conditions commonly
used in quantum MC simulations (alternatively, open or cylindrical boundaries that are more practical with some other methods \cite{stoudenmire12})
are not compatible with the state-operator correspondence and, therefore, the level spectrum of the Hamiltonian defined on a torus does
not correspond to the spectrum of CFT scaling dimensions in a known way \cite{schuler16}. The proper conformal spatial geometry is the surface
of a sphere \cite{cardy85}, which is very difficult to realize uniformly for lattice models \cite{brower18,brower21,lao23,ayyar23,brower24}. Here
the fuzzy sphere approach is making inroads \cite{zhu23,hoffmann23,zhou24,chen24a,chen24b,fardelli24}, with the spherical geometry realized with fermions
coupled to a magnetic monopole \cite{haldane83} and interactions designed to implement desired symmetries and quantum phase transitions when projected
to the lowest Landau level. Though remarkable results have already been presented for some models, there are still significant limitations in
the system size (number of fermions), and it is also often not obvious how to construct a model with a specific type of quantum phase transition.

The covariance method presented here does not rely on the specific conformal geometry, as will be explicitly demonstrated numerically in the later
sections by classical MC results generated on standard periodic square and cubic lattices of volume $L^2$ and $L^3$, respectively instead of the
infinite cylinders in $d=2$ or radial quantization in $d=3$. From the CFT (or other field theory) perspective,
it is from the outset not clear, however, whether diagonalizing
a covariance matrix Eq.~(\ref{cij}) will truly yield eigenvalues decaying asymptotically as $r^{-2\Delta_i}$ with (a) different scaling dimensions
$\Delta_i$ and, related to this, (b) not including any off-diagonal $r^{-(\Delta_i+\Delta_j)}$ contributions. These complications might be expected on account of
the fact that the CFT operators belonging to the same multiplet $p$ are not orthogonal in the sense of correlation functions. It will be argued below that
diagonalization of an $N \times N$ covariance matrix will nevertheless produce eigenvalues decaying asymptotically as $r^{-2\Delta_i}$, with the $N$ smallest
scaling dimensions $\Delta_i$ for a system described by a CFT. In practice, statistical errors in MC simulations will limit the number of such decay
exponents that can be observed in practice, as will be seen in the later sections. However, even a rather small number of scaling dimensions (in
each symmetry sector) beyond the normally accessible lowest primary can still provide very valuable information when probing a system for
its CFT description.

\subsection{Covariance matrices of ideal lattice operators}
\label{sec:cftcov}

In principle, it is possible to construct lattice operators in a critical system that only contain a single primary or descendant CFT operator,
though such operators would normally be extremely complicated and impossible to design in practice. Considering an infinite set $Q_i({\bf r})$ of
these ideal operators in a specific symmetry sector, including a fixed Lorentz spin $l$, arranged as a column vector $Q({\bf r})$;
\begin{equation} \label{opvector}
Q^T({\bf r}) = [Q_0({\bf r}),Q_1({\bf r}),\ldots ].
\end{equation}  
We will first assume that this group of operators only includes one primary $Q_0$, with scaling dimension $\Delta_0$, along with all its descendants
$Q_n$, $n=1,2,\ldots$, within the same symmetry sector. Then
\begin{equation} \label{delta_n}
\Delta_n = \Delta_0 + 2n,
\end{equation}  
since, for a primary operator with given $l$, the descendants with scaling dimensions $\Delta_0 + 2n-1$ belong to different spin
sectors; $l \pm 1$. The second descendant is therefore the lowest one with the same $l$ as the primary.

With the restriction to a single conformal multiplet, none of the operators are orthogonal to each other in the sense of the correlation
functions (only operators in different multiplets are), i.e.,
\begin{equation}\label{cij}
C_{ij}({\bf r}) = \langle Q_i({\bf r})Q_j({\bf 0})\rangle = c_{ij} r^{-(\Delta_i+\Delta_j)},
\end{equation}
where the factors $c_{ij}$ should be expected to be nonzero for all $i,j$. These correlation functions form the covariance matrix $C$, which can be
written as
\begin{equation}  
 C({\bf r}) = \langle Q({\bf r})Q^T({\bf 0})\rangle.
\end{equation}  
This matrix can be diagonalized by a unitary transformation $U$ that depends on the separation ${\bf r}$,
\begin{eqnarray}
  D({\bf r}) & = & U^T({\bf r})\langle Q({\bf r})Q^T({\bf 0})\rangle U({\bf r}) \nonumber  \\
    & = & \langle U^T({\bf r})Q({\bf r})Q^T({\bf 0})U({\bf r})\rangle   \nonumber  \\
    & = & \langle \tilde Q({\bf r})\tilde Q^T({\bf 0})\rangle, 
\label{dwqtilde}
\end{eqnarray}
with $D_{ij}({\bf r}) = 0$ for $i \not=j$ and the transformed operators $\tilde Q({\bf r})=U^T({\bf r})Q({\bf r})$ have been defined.

The question now is whether the eigenvalues of $C$ decay according to the corresponding scaling dimensions of the original operators, i.e.,
whether is it true that
\begin{equation} \label{dii}
D_{i}({\bf r}) \equiv D_{ii}({\bf r}) = A_i r^{-2\Delta_i},
\end{equation}
where the prefactors $A_i$ should be independent of $r$. These decay forms, purely governed by the individual exponents $2\Delta_i$, are not obvious even
if $D$ is diagonal, since the transformation might in principle mix the different power laws appearing in the original covariance matrix $C$. While a
completely rigorous proof of Eq.~(\ref{dii}) will not be presented here, a series of plausible arguments will be put forward in its support.
Thus assuming that Eq.~(\ref{dii}) holds, subsequently Sec.~\ref{sec:lattcov} will proceed with a demonstration of the disentangling of scaling
dimensions by diagonalization of a covariance matrix computed with a set of simple lattice operators that can be designed in practice.

Consider the elements $u_{ij}$ of a matrix $u$ that defines a transformation for a finite set of uperators; the vector in Eq.~(\ref{opvector})
truncated at some $n=m$ that later can be taken to infinity;
\begin{equation}
\tilde Q_n = \sum_{i=0}^m u_{ni}Q_i,~~~n=0,1,\ldots,m,
\label{tildeqn}
\end{equation}
where initially $u$ may be different from the diagonalizing transformation $U({\bf r})$ but $u \equiv U({\bf r})$ will be established along with
the separation of eigenvalues according to Eq.~(\ref{dii}). The transformed covariance matrix is 
\begin{equation} \label{dii2}
D_{ij}({\bf r}) = \sum_{kp} u_{ik}u_{jp}\langle Q_kQ_p\rangle = \sum_{kp} u_{ik}u_{jp}C_{kp}({\bf r}), 
\end{equation}
which is not yet assumed to be diagonal. Using the covariance matrix in the form of Eq.~(\ref{cij}) and the descendant dimensions
in Eq.~(\ref{delta_n}), we can also write $D({\bf r})$ as
\begin{equation} \label{dii3}  
D_{ij}({\bf r}) = r^{-(\Delta_i+\Delta_j)} \sum_{kp} u_{ik}u_{jp}c_{kp}r^{-2(k+p-i-j)}, 
\end{equation}
so that the original covariance matrix Eq.~(\ref{cij}) is obviously recovered when $u_{ik}=\delta_{ik}$ and $u_{jp}=\delta_{jp}$.

If we require that the transformation leads to a diagonal $D({\bf r})$,
the eigenvectors $u^T_i = [u_{i0},\ldots,u_{iN}]$ clearly must depend on $r$. However, we can make the plausible assumption that a collective
solution for all ${\bf r}$ must exist, so that a single matrix diagonalization is sufficient for constructing a diagonal $D({\bf r})$ for all ${\bf r}$.
This collective solution can be found by assuming that the $r$ dependence of $u$ completely cancels $r^{-2(k+p-i-j)}$ in Eq.~(\ref{dii3}), i.e., defining
a matrix $v$ with
\begin{equation} \label{uv}
v_{ik} \equiv u_{ik}r^{-2(k-i)},
\end{equation}
and positing that this matrix can now be independent of $r$. Eq.~(\ref{dii3}) then becomes
\begin{equation} \label{dii4}
D_{ij}({\bf r}) = r^{-(\Delta_i+\Delta_j)} \sum_{kp} v_{ik}v_{jp}c_{kp}, 
\end{equation}
which represents a standard orthogonal transformation of the symmetric matrix $c$. For any such transformation, the general original form Eq.~(\ref{cij}) is
maintained and the special unique transformation that diagonalizes $c$ leads to eigenvalue separation according to Eq.~(\ref{dii}), with the factors
given by
\begin{equation} \label{ai}
A_i=\sum_{kp} v_{ik}v_{ip}c_{kp}. 
\end{equation}
It should be stressed again that the only assumption made here was that a collective solution for all ${\bf r}$ exists. The specific $r$
dependent prefactor $r^{-(\Delta_i + \Delta_j)}$ in front of the sum representing the collective transformation in Eqs.~(\ref{dii3}) and (\ref{dii4}) is
then required in order for the sum to contain the powers of $r$ present originally and, eventually, to obtain the correct set of $m+1$ scaling dimensions.
Finally we can also let $m \to \infty$ in Eq.~(\ref{tildeqn}).

\subsection{Covariance of simple lattice operators}
\label{sec:lattcov}

It is in general not possible in practice to uniquely define a lattice operator that contains a single field operator. Explicit translations from the lattice to
the field operators are only known in special cases \cite{affleck87,yellowbook,patil18} and typical operators defined for generic lattice models contain all the
symmetry compatible CFT operators. Using a finite set of $N$ lattice operators, it should be possible, however, to find optimal linear combinations of these
operators such that their correlation functions exhibit asymptotic decays according to the $N$ lowest scaling dimensions. While these optimal operators of course
depend on the original set of $N$ chosen operators, in general one would expect some overlap between arbitrary lattice operators and the field operators.

For a given set of symmetries, many conformal multiplets with their full towers will in general contribute to a simple lattice operator $O_i$, which therefore
be written as a linear combination of the idealized operators $Q_{pi}$ belonging to several multiplets $p$ and $i=0,1,\ldots,\infty$. Alternatively, we can
expand in the correlation orthogonal operators $\tilde Q_{pi}$ discussed above in Sec.~\ref{sec:cftcov}, which is what we will do here. For the sake of a
simpler notation, we will just use a collective index $k=1,2,\ldots, \infty$ to refer to all of these, since operators in different multiplets are
automatically correlation orthogonal. Thus, our set of $N$ lattice operators can in principle be written as 
\begin{equation} \label{lattop1}
O_i = \sum_{k=1}^\infty f_{ik} \tilde Q_{k},~~~i=1,\ldots,N,
\end{equation}
where the exact form of the transformed field operators $\{\tilde Q_{k}\}$ defined by Eq.~(\ref{dwqtilde}) depends on the distance between operators that
eventually will be used in correlation functions, as detailed in Sec.~\ref{sec:cftcov}. Without explicitly indicating this $r$ dependence, the covariance
matrix of the lattice operators is
\begin{equation} \label{lcov1}
C_{ij}({\bf r}) = \sum_{k=1}^\infty\sum_{p=1}^\infty f_{ik}f_{jp} \langle \tilde Q_{k}({\bf r}) \tilde Q_{p}({\bf 0})\rangle,
\end{equation}
where the elements of $f$ may also have some (weak) $r$ dependence, as will be further discussed below.
Since the $\tilde Q$ operators diagonalize the covariance matrix for given $r$, with eigenvalues given by Eq.~(\ref{dii}), we have
\begin{equation} \label{lcov2}
C_{ij}({\bf r}) = \sum_{k=1}^\infty f_{ik}f_{jk}A_kr^{-2\Delta_k}.
\end{equation}

Diagonalizing $C$ corresponds to finding specific linear combinations of the $N$ lattice operators,
\begin{equation} \label{lattop2}
\tilde O_i = \sum_{a=1}^N g_{ia} O_{a},~~~i=1,\ldots,N,
\end{equation}
and we again denote by $D({\bf r})$ the result of the orthogonal transformation of $C({\bf r})$ that will eventually be diagonal;
\begin{equation} \label{lcov3}
D_{ij}({\bf r}) = \sum_{a=1}^N\sum_{b=1}^N g_{ia}g_{jb} C_{ab}({\bf r}),
\end{equation}
which with Eq.~(\ref{lcov2}) becomes
\begin{equation} \label{lcov4}
D_{ij}({\bf r}) = \sum_{a=1}^N\sum_{b=1}^N\sum_{k=1}^\infty g_{ia}g_{jb}f_{ak}f_{bk}A_kr^{-2\Delta_k}.
\end{equation}
Here the sums over $a$ and $b$ just represent transformed coefficients defining a matrix $w$; thus
\begin{equation} \label{lcov5}
D_{ij}({\bf r}) = \sum_{k=1}^\infty w_{ik}w_{jk}A_kr^{-2\Delta_k}.
\end{equation}
If the sum over $k$ would extend only to $k=N$, we could now simply conclude that a diagonaliing transformation corresponds to $w_{ik}=\delta_{ik}$
and a complete separation of the different scaling dimensions, with the eigenvalues $D_{i}$ then decaying as in Eq.~(\ref{dii}). The fact that the
sum includes an infinite number of power laws implies that the decoupling is not perfect. However, with the rapidly increasing exponents $2\Delta_k$
with $k$, the remaining mixing should be completely negligible even when $N$ is not very large, as we will indeed see in the numerical examples
in the following sections.

Note that, while it is only the elements of $g$ in Eq.~(\ref{lcov3}) that are determined by diagonalization of a numerically computed covariance matrix
$C$, in principle the $f$ coefficients in the expansion of the lattice operator into the correlation orthogonal field operators can also be determined
from $g$, the trivial (unit) matrix $w$ (under the assumption of no contributions from scaling dimensions $\Delta_k$ with $k>N$), and the prefactors
$A_k$ of the power law decays in Eq.~(\ref{lcov5}). The expansion in the conventional operators $Q$ requires additional CFT information in the form
of the coefficients $c_{ij}$ in Eq.~(\ref{cij}). We will here not be concerned with these potential further useful aspects of the covariance method
beyond the eigenvalues.

Unlike the ideal operators (with a single scaling dimension) considered in Sec.~\ref{sec:cftcov}, the complete disentanglement in the exact form of
Eq.~(\ref{dii}) should only apply asymptotically for large $r$. Moreover, a critical point of a lattice model does not correspond exactly to the CFT fixed
point but only flows to this point when the
lattice size $L$ and the distance $r$ (in correlation functions) are taken to infinity. For finite length scales, irrelevant perturbations are present and
lead to multiplicative corrections of the form $(1+\sum_i\lambda_i L^{-\omega_i})$ or $(1+\sum_i\lambda_i r^{-\omega_i})$, where $\omega_i = \Delta_i-d$ for
all symmetry allowed operators with $\Delta_i > d$ (and the relevant perturbations with $\Delta_i < d$ have been tuned away to reach the critical or
multi-critical point). These corrections are well known for critical scaling of observables in finite lattices \cite{fisher72,binder81,barber83} and
should also affect the eigenvalues of the covariance matrix. The corrections should stem from the expansion of the lattice operators in the field
operators in Eq.~(\ref{lcov1}), where the coefficients $f$ would exhibit some $L$ and $r$ dependence. However, asymptotically the unknonw matrix $f$ and the
diagonalizing transformation $g$ in Eq.~(\ref{lcov3}) should be independent of $L$ and $r$.

The finite number $N$ of operators is also of course a limitation, as already discussed above, but still one would expect to resolve $N$ scaling dimensions
in principle. However, numerical errors in the original covariance matrices $C({\bf r})$, in particular statistical MC errors, will in practice make it
impossible to determine more than a small number of scaling dimensions, considering the rapid increase in the decay power (in steps of $4$) with the
descendant index $n$, Eq.~(\ref{delta_n}), for operators with fixed Lorentz spin $l$. The results presented below will show that one or two descendant
dimensions can typically be extracted along with the primary, which is still very useful and exceeds what can be achieved by fitting conventional
correlation functions.

\section{2D Ising model} \label{sec:is}

As a first test, we here study the conventional classical Ising model with Hamiltonian
\begin{equation}
H = - \sum_{\langle ij\rangle}\sigma_i\sigma_j,~~\sigma_i \in \{-1,+1\},
\label{isham}
\end{equation}
with the spins residing on an $L\times L$ square lattice with standard periodic boundary conditions. With the coupling constant set to unity above, the
critical temperature is $T_c = 2/\ln[1+\sqrt{2}]$ and we study the model with MC simulations at this point, using a combination of Swendsen-Wang
\cite{swendsen87} and Wolff \cite{wolff89} cluster updates, the latter with the number of clusters constructed in each MC step adapted so that roughly
$N$ spins are visited. The covariance matrices of all the operators illustrated in Fig.~\ref{ops_is} are computed for ${\bf r}=(r,0)$ and $(0,r)$ (and
these are averaged) after every few MC steps and binned, with a data bin representing averages over $10^6$ or more MC steps. Diagonalization is
carried out post simulation using a large number (of the order $10^4$) of stored data bins, using bootstrap analysis for statistical errors.

A full symmetrized operator $O_i$ corresponds to starting from a single operator product $O_{ia}$, $a=1$, inside the $3\times 3$ cell as in
Fig.~\ref{ops_is}, which can always be written as
\begin{equation}
O_{ia}({\bf r}) = 
\prod_{\{c({\bf r})\}}\rho_c,~~ \rho_c \in \{\sigma_c,I\}, 
\end{equation}
where $\{c({\bf r})\}$ is the set of $9$ sites belonging to the cell centered at ${\bf r}$ and $\rho_c$ is either an Ising spin operator (for the sites
with a red dot in Fig.~\ref{ops_is}) or the identity operator $I$ (sites with no dot). Symmetrization corresponds to summing over all unique symmetry
equivalent locations of the $\sigma_c$ products, generated by rotating about the cell center and reflecting about the central horizontal or vertical axis. Thus,
the symmetrized operators can be expressed as
\begin{equation}
O_i = \frac{1}{n_i}\sum_{a=1}^{n_i}O_{ia}({\bf r}),
\end{equation}
where $a$ refers to the different transformed operator patterns and $n_i$ is the number of such unique patterns, which is indicated for each operator
in Fig.~\ref{ops_is}. The distance $r$ between operators in a correlation function Eq.~(\ref{cij}) is defined as the center-center distance between the
$3\times 3$ spin cells. We will first study the $r$ dependent eigenvalues for a fixed system size, mainly $L=256$ but with some results also for $L=128$
to check for finite-size effects. After systematically investigating the role of the number $N$ of operators, we we will also consider smaller systems
and study the covariance matrices for $r=L/2$ versus $L$.

The goal here is to resolve some of the known scaling dimensions of the 2D Ising critical point. For 2D CFTs with central charge $c \le 1$, the possible
values of $c$ are $c=1-6/[m(m+1)]$ for integer $m\ge 3$. For given $m$, the primary scaling dimensions take the form \cite{yellowbook}
\begin{equation} \label{cftc1}
\Delta_{jk}=\frac{[k+m(k-j)]^2-1}{2m(m+1)},~~1 \le j \le k < m.
\end{equation}
The case $m=3$ ($c=1/2$) corresponds to the conventional Ising critical point, while $m=4$ ($c=7/10$) is the tricritical Ising class that will be studied with
the BC model in Sec.~\ref{sec:bc}. The two $l=0$ primaries for $m=3$ (corresponding to the spatial symmetrized operators described above) are $\Delta_{22}=1/8$
($Z_2$ odd) and $\Delta_{21}=1$ ($Z_2$ even), which for simplicity we will just refer to as $\Delta_0$ in their respective $Z_2$ sectors.

\begin{figure}[t]
\includegraphics[width=82mm]{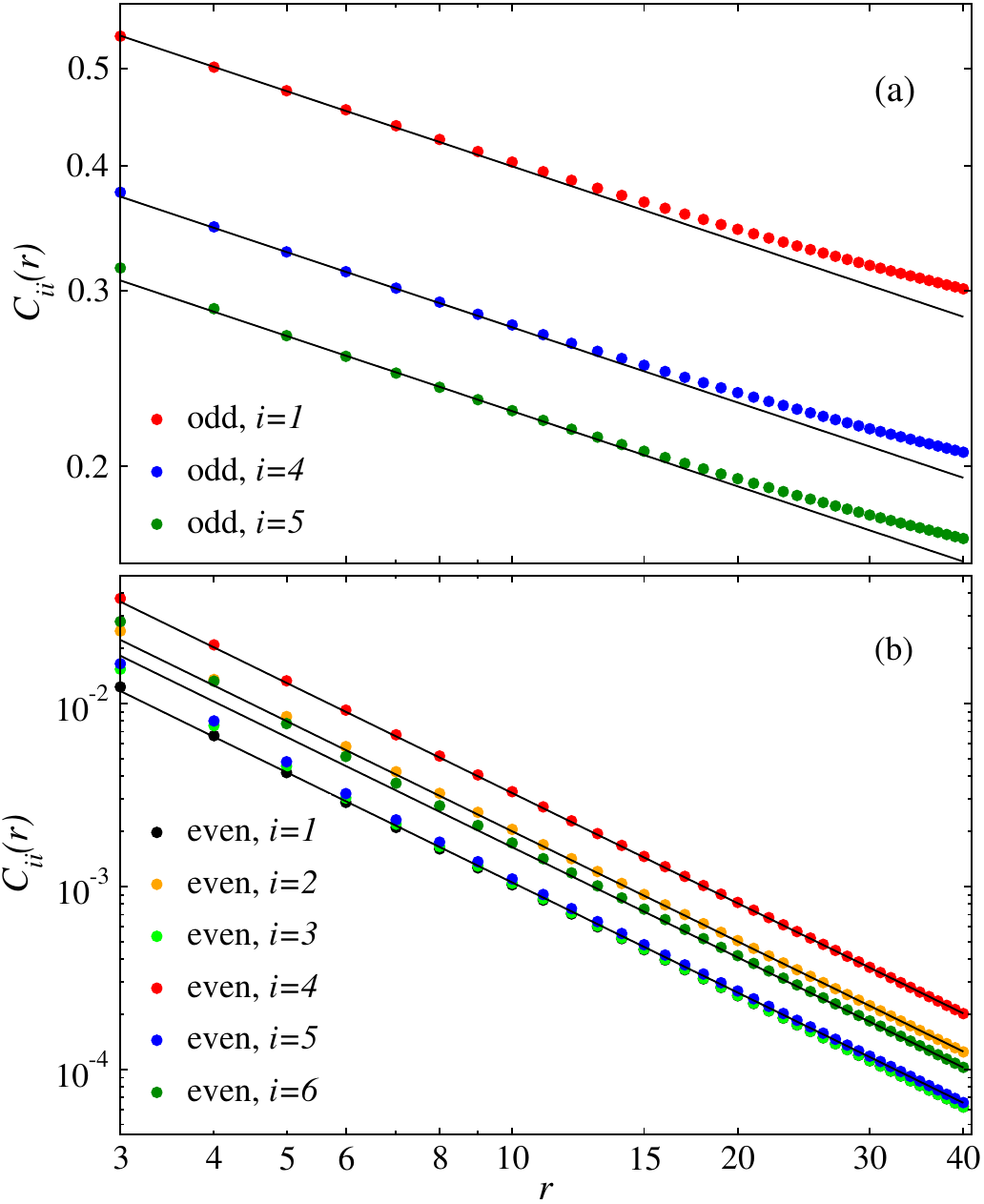}
\caption{Diagonal elements of covariance matrices for $Z_2$ odd (a) and even (b) operators, shown in log-log plots for lattice size $L=256$ vs $r \ll L$.
The index $i$ corresponds to Fig.~\ref{ops_is}. In (a) the $i=2$ and $i=3$ correlations coincide very closely with those for $i=1$ and are
not shown. The $i=6$ correlations are very close to those for $i=4$ and are also not shown. The lines in (a) show the behavior $\propto r^{-1/4}$
expected for large $r$ and $L$ with $r \ll L$, and the lines in (b) similarly show the expected form $\propto r^{-2}$.}
\label{oicor}
\end{figure}

\begin{figure*}[t!]
\includegraphics[height=90mm]{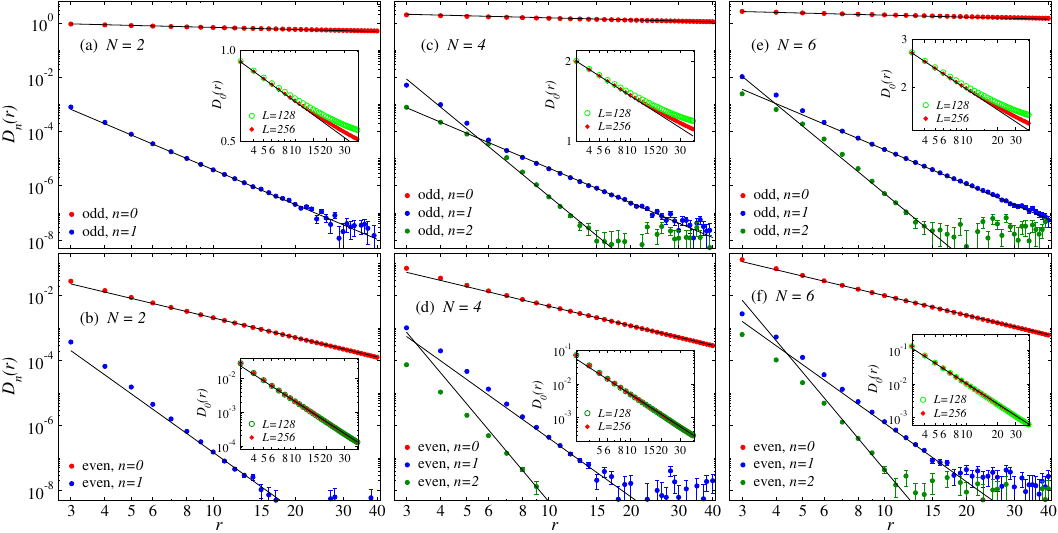}
\caption{Distance dependent eigenvalues of covariance matrices involving $N=2$ Ising operators in (a) and (b), $N=4$ in (c) and (d), and
$N=6$ in (e) and (f). In the notation of Fig.~\ref{ops_is}, the set of antisymmetric operators is $\{O_1,O_4\}$ in (a), $\{O_1,\ldots,O_4\}$ in (c),
and $\{O_1,\ldots,O_6\}$ in (e). The symmetric operators are $\{O_1,O_5\}$ in (b), $\{O_1,O_4,O_5,O_6\}$ in (d), and $\{O_1,\ldots,O_6\}$ in (f).
In the main graphs the lattice size is $L=256$ and the insets show $L=128$ and $L=256$ results for the largest eigenvalue in each case. The
lines in these log-log plots show the expected power-law decay $\propto r^{-2\Delta_0}$, corresponding to the CFT primary operator with $\Delta_0=1/8$
in the top row and $\Delta_0=1$ in the bottom row, as well as the faster decaying power-laws originating from the leading descendants,
$\propto r^{-2(\Delta_0+2)}$ and, in the case of (c)-(f), also $\propto r^{-2(\Delta_0+4)}$.}
\label{n246}
\end{figure*}

\subsection{Single-operator correlations}
\label{sec:singleops}

The correlation functions of the six individual operators in both symmetry sectors, i.e., the diagonal elements of the covariance matrix $C_{ij}(r)$,
are graphed versus $r$ in log-log plots in Fig.~\ref{oicor}. Given that all the lattice operators should contain the primary CFT operator
in the respective $Z_2$ sectors, asymptotic decays of the form $r^{-2\Delta_0}$ are expected, with $\Delta_0=1/8$ in the odd sector and $\Delta_0=1$
in the even sector. The small value of the odd scaling dimension implies larger effects of the periodic boundaries, i.e., the $1 \ll r \ll L$
limiting behavior is harder to realize, even on lattices much larger than $L=256$ used here. Indeed, clear deviations from the expected behavior are
seen at the larger values of $r$ in Fig.~\ref{oicor}(a), while in the small range $3 < r < 10$ the expected behavior is reasonably well reproduced.
In Fig.~\ref{oicor}(b), the decay exponent is much larger and no boundary enhancements of the correlations are detectable within statistical errors.

In principle, it should be possible to gain some
information also on the descendant scaling dimensions from these individual correlation functions, at least those in Fig.~\ref{oicor}(b). However, as
already discussed in Sec.~\ref{sec:lattcov}, the short distance correlations not only mix contributions with different decays $r^{-2\Delta_n}$ but also
contain other (related) power laws originating from various scaling corrections. It is therefore not possible in practice to reliably extract even
the first visible descendant dimension $\Delta_1=\Delta_0+4$, in particular in the odd $Z_2$ sector, Fig.~\ref{oicor}(a), where also the boundary effects
are very significant.

\subsection{Covariance eigenvalues for large $L$ and $r \ll L$}
\label{sec:islarge}

In the MC simulations all correlations were computed between $N=6$ operators within the even and odd sets illustrated in Fig.~\ref{ops_is}, i.e.,
in each case $36$ different correlation functions. Any subset of these correlations forming
a smaller matrix can be selected post simulation for the diagonalization process. When selecting such subsets of operators (and of course also
when constructing the full set), there are opportunities for optimization, and some results showing variations in results for different subsets will
be discussed below. First, Fig.~\ref{n246} shows $L=256$ results for eigenvalues of covariance matrices computed with $N=2$, $4$ and $6$ operators,
with the operators included for $N=2$ and $N=4$ not necessarily the optimal ones but among the best ones from the full set of size operators. Here we
also refer to the eigenvalues $D_n(r)$ for given $r$ using the index $n=0,\ldots,N-1$ from the smallest to the largest. Then the large-$r$ decay form
should match the scaling dimension with the same label; $D_n \sim r^{-2\Delta_n}$. However, in some cases the order of the
eigenvalues is different for small $r$.

For $N=2$, Figs.~\ref{n246}(a) and \ref{n246}(b), the expected power law decays corresponding to the two lowest scaling dimensions can be resolved, as
demonstrated  with the straight lines showing the expected decay forms. There are some deviations from these ideal forms at short distances, on account of
the corrections discussed in Sec.~\ref{sec:lattcov}. In the case of the largest eigenvalues in the odd sector, Fig.~\ref{n246}(a), there are also large
deviations at the
longer distances, as shown in more detail in the insets for both $L=128$ and $L=256$. The upward deviations from the power law decay, with larger deviations
for the smaller system size, are clearly again due to effects of the periodic boundary conditions, as in Fig.~\ref{oicor}(a). It is noteworthy, however,
that the enhancement of the correlations for $L=256$ are actually a bit smaller in the eigenvalue than in the diagonal elements $C_{ii}$, for reasons
that are not obvious.

The reason for the overall significant boundary effects in Figs.~\ref{oicor}(a) and \ref{n246}(a), i.e., the difficulty in realizing the $1 \ll r \ll L$ limit,
stems from the small value of the scaling dimension of the odd primary, the decay exponent being $2\Delta_0=1/4$. There are no apparent boundary effects
in the fast decaying second eigenvalue in Figs.~\ref{n246}(a), though the error bars are large for $r \agt 20$ and might in principle hide a small upward deviation.
Significant boundary effects are unlikely, however, because the expected (and numerically confirmed) decay exponent is large, $2\Delta_1=4.25$,
and the relative size of the boundary correction should be of order $[r/(L-r)]^{-2\Delta_1}$. In the even sector, Fig.~\ref{n246}(b), the exponent
of the largest eigenvalue is $2\Delta_0=2$, and also here there are no discernible boundary effects or differences between $L=128$ and $256$ for the distances
$r \le 40$ graphed, and no upward deviations are seen at the largest $r$. In this case, the second eigenvalue is affected by larger relative corrections
at short distances, and the expected exponent $2\Delta_1=6$ can be observed only rather close to the values of $r$ at which the error bars begin to render
the data useless for fitting.

Moving to the $N=4$ results in Figs.~\ref{n246}(c) and \ref{n246}(d), here also the second descendants can be resolved, though marginally
in the case of the even sector, where only the points $r=7,8,9$ fall close to the expected form before the error bars become too large.
The fourth eigenvalues are completely noise dominated overall and are therefore not shown. In Fig.~\ref{n246}(c) the $n=1$ and $n=2$
eigenvalues cross each other close to $r=6$. The color coding of the data points only refer to the ordering of the eigenvalues for each $r$, but
it is clear from the crossing behavior that the association of the data points with scaling dimensions swap. This behavior is confirmed by examining
the eigenvectors, the elements of which are graphed versus $r$ in Fig.~\ref{vec}. All the $n=1$ and $n=2$ elements swap between $r=6$ and $r=7$,
and then become essentially $r$ independent for $r > 8$. Note that all four operators contribute to the diagonalizing linear combinations.

Increasing the number of operators to $N=6$, Figs.~\ref{n246}(e) and \ref{n246}(f) still only include three eigenvalues because the higher ones
are again completely noise dominated. The results qualitatively do not look very different from those obtained with $N=4$, but it should be
noted that the overall amplitudes of the eigenvalues increase somewhat with $N$. While the impact of the statistical noise also increases slightly,
the relative errors overall diminish with increasing $N$---this trend is clear at least for the $n=1$ eigenvalues. Thus, using larger $N$ may
be helpful even if the number of resolved scaling dimensions is less than $N$. The favorable effect of the overall scale of the eigenvalues
increasing with $N$ likely reflects improving flexibility in forming linear combination of the lattice operators that maximize the overlaps with
the field operators. However, it should be kept in mind that the computational effort of accumulating the covariance matrix scales as $N^2$ and
can easily dominate the total effort of simulations. Realistically, $N$ would therefore typically be no larger than five to ten.\footnote{In principle,
two-point and multi-point correlation functions could be computed at arbitrary separation and an essentially unlimited range of covariance matrices
could be constructed with them post simulation. In cases such as the simple Ising models, these correlation functions can also be computed using
improved cluster estimators to reduce statistical errors and improve computational efficiency. Here the specific lattice operators were only
implemented with standard estimators by simply evaluating the corresponding cell values for each spin configuration and accumulating the relevant
products of those numbers.}

\subsection{Statistical errors}
\label{sec:errors}

It should be noted here that the noise level can be much smaller than what would be expected from the individual correlation functions $C_{ij}(r)$,
such as the diagonal ones in Fig.~\ref{oicor}, where the statistical errors of the $Z_2$-odd operators are of the order $10^{-6}$ or above, while
in Fig.~\ref{n246} the eigenvalues are resolved down to values as small as $10^{-8}$. The reason why the diagonalizaton procedure can resolve such small
eigenvalues must be that the MC statistical fluctuations of $C_{ij}$ for all $i,j$ are correlated, i.e., the individual error bars do not give a complete
picture of the information contained in the entire data set. This effect of noise correlations is less pronounced for the $Z_2$-even operators, where the
individual correlation functions also have statistical errors of order $10^{-8}$, similar to the noise cut-offs in Fig.~\ref{n246}.

\begin{figure}[t]
\includegraphics[width=80mm]{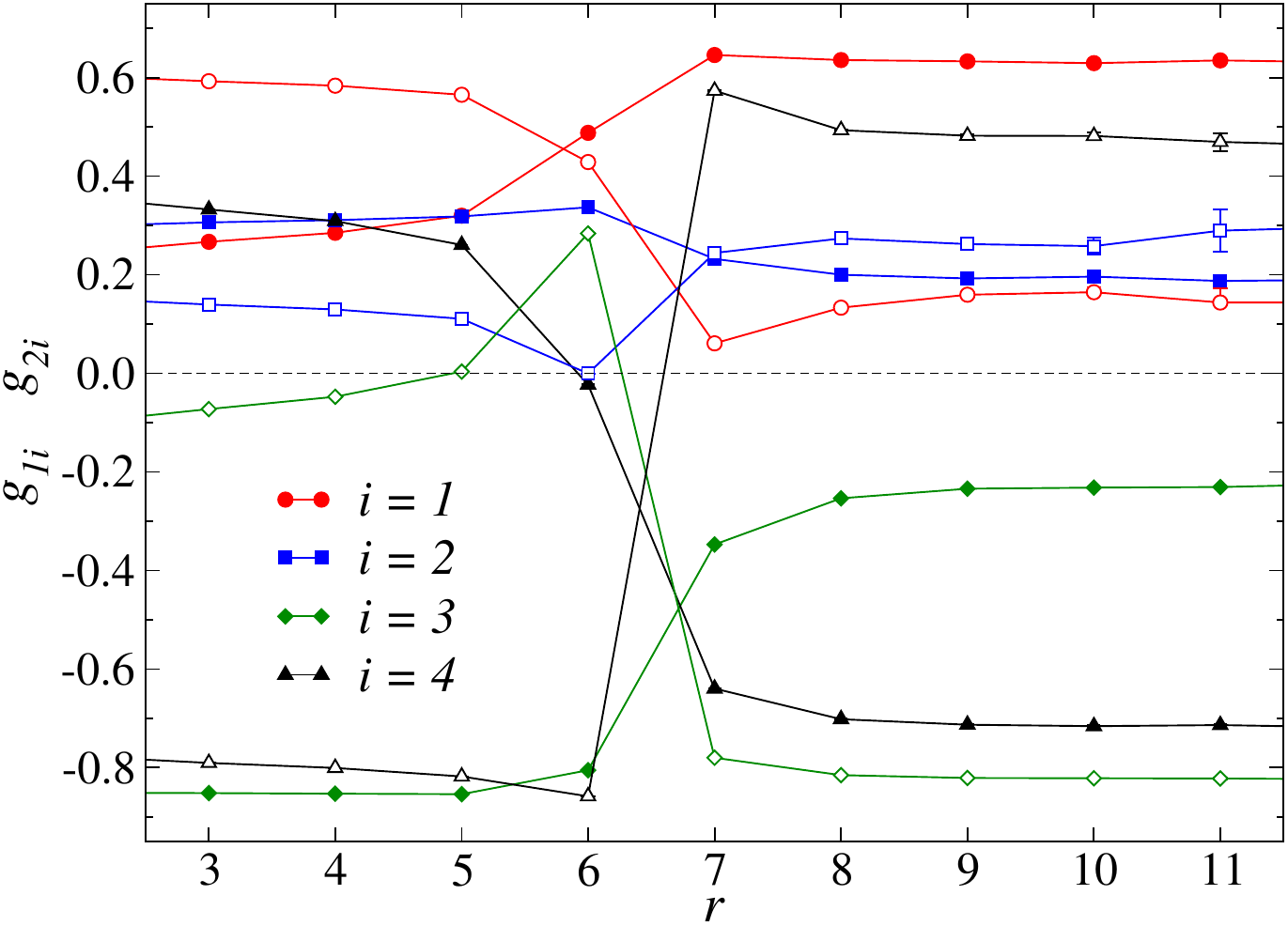}
\caption{Distance dependent elements of the eigenvectors corresponding to the $n=1$ and $n=2$ eigenvalues in Fig.~\ref{n246}(c). The filled symbols
of different colors show the $n=1$ coefficients $g_{1i}$ and the corresponding $n=2$ values $g_{2i}$ are shown with the open symbols of matching
shapes and colors.}
\label{vec}
\end{figure}

At face value, it may appear as if many eigenvalues in Fig.~\ref{n246} for large $r$ are stable (with relatively small error bars) around values of order
$10^{-8}$, above the expected much smaller values. However, this apparent flaw is mainly a quirk of the logarithmic scale when graphing errors defined
as one standard deviation of the mean. As an example, the $n=1$ eigenvalue at $r=29$ in Fig.~\ref{n246} is less than $2.5$ times its own error bar, thus
reasonably consistent with an extrapolated value $\approx 2\times 10^{-9}$ if the power law decay for the smaller distances continues to apply as expected.
There may still be a small positive bias on the overall errors. Statistically, a roughly equal number of positive and negative eigenvalues should be expected
when the actual eigenvalues are extremely small. However, there are no negative $n=1$ eigenvalues in Fig.~\ref{n246}(f) [while there are some in the
$n=2$ set and also other graphs, except for in Fig.~\ref{n246}(a)].

A small positive bias in the eigenvalues could possibly reflect the fact that matrix diagonalization is a nonlinear procedure, which implies that the statistical
errors of the eigenvalues do not have zero expectation value for a finite data set. As a trivial example of such noise bias, consider an MC estimate
$\bar A$ of some expectation value $\langle A\rangle$, which can be written as $\bar A = \langle A\rangle +\delta$, where the unknown error $\delta$ is of
order $M^{-1/2}$, where $M$ is the total number of MC measurements performed. Denoting an expectation value over different MC simulations by $[]$, we have
$[\bar A] = \langle A\rangle$ because $[\delta]=0$. However, for the square of the MC estimate
we have $[\bar A^2] = \langle A\rangle^2 + [\delta^2]$, i.e., it is biased toward higher values. The bias vanishes at the rate $M^{-1}$ but can be statistically
significant even for large $M$ if $\langle A\rangle$ is very small or if the prefactor is large. In the case of matrix diagonalization, a quadratic
error leading to an $M^{-1}$ bias can similarly be expected in the eigenvalues (which can easily be confirmed for a $2\times 2$ matrix), and it can also
be expected that the prefactor of the $M^{-1}$ scaling form increases with $N$.

\begin{figure}[t]
\includegraphics[width=80mm]{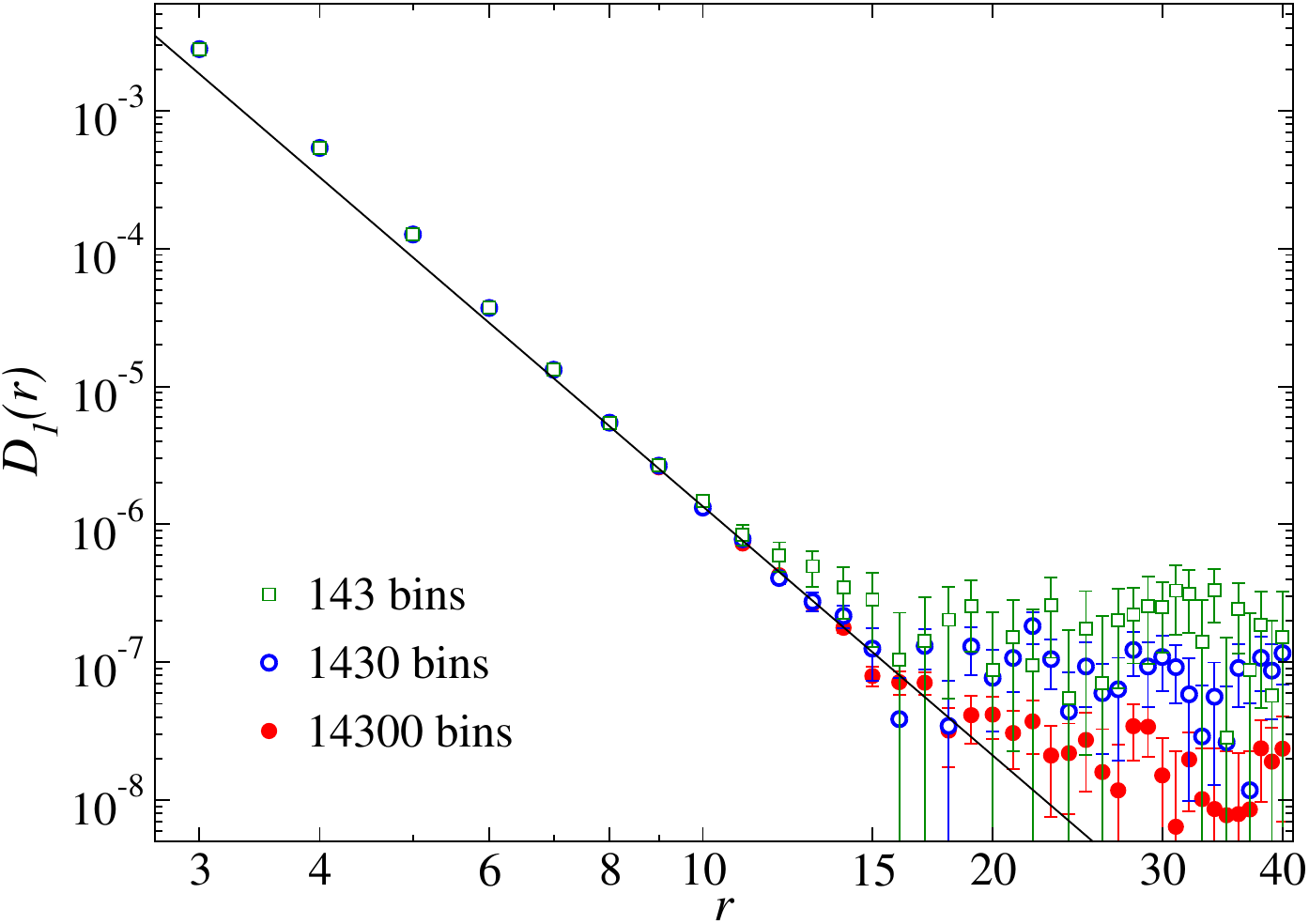}
\caption{The $n=1$ eigenvalues from Fig.~\ref{n246}(f) (red solid circles) graphed along with results obtained when using only $10\%$ (blue open circles)
and $1\%$ (green open squares) of the available 14300 data bins. Each bin corresponds to $10^6$ MC steps. The line shows the decay form $r^{-6}$
expected with the $n=1$ scaling dimension $\Delta_1=3$.}
\label{less}
\end{figure}

To test the ability to resolve the eigenvalues with data noisier than in Fig.~\ref{n246}, in Fig.~\ref{less} the $n=1$ eigenvalues from Fig.~\ref{n246}(f)
are graphed along with results obtained when only a small fraction, $10\%$ and $1\%$, of the original data set is used. When the size of the data set is reduced,
the positive values of the noise dominated large-$r$ eigenvalues increase overall roughly in the way statistically expected, i.e., with $1\%$ of the
data used, the noise level (taken as the standard deviation or typical magnitude of the positive eigenvalues for large $r$) increases by an order of magnitude.
Again, the logarithmic scale can be misleading for the error bars, as the largest deviation from $0$ in the $1\%$ data set is about $2.5$ error bars. There are no
negative mean values in the data sets in Fig.~\ref{less}, suggesting a positive bias. The bias should grow by a factor of $100$ when reducing the data set to
$1\%$, which is not seen in Fig.~\ref{less}, though the increase may be marginally above the factor $10$ expected for unbiased statistical fluctuations. The
bias effect, quantified by the unknown prefactor of the $M^{-1}$ scaling of the quadratic error discussed above, therefore must be very small, though it may
still be responsible for the lack of negative eigenvalues for large $r$, in combination with the fact that the true eigenvalues are positive though
very small. The positive bias may be of significance in the way the eigenvalues initially begin to deviate from the power law decay, i.e., above $r \approx 10$
in the $1\%$ data but only above $r\approx 20$ when the full data set is used. The deviations are mainly upward, a behavior typically (not always) seen
before the data become completely noise dominated with the apparent flattening-out behavior discussed above.

As an example of the computational effort spent on these calculations, in the case of the $L=256$ lattice (all the data in Figs.~\ref{n246} - \ref{less})
each data bin consisted of $10^6$ MC steps and the covariance matrices were accumulated after every $10$ steps. Each bin required about 10 CPU (core) hours of
compute time and the total of more than $10^4$ bins were generated by running independent simulations simultaneously on about $100$ cores.

\subsection{Dependence on the operator set}
\label{sec:operators2}

\begin{figure}[t]
\includegraphics[width=82mm]{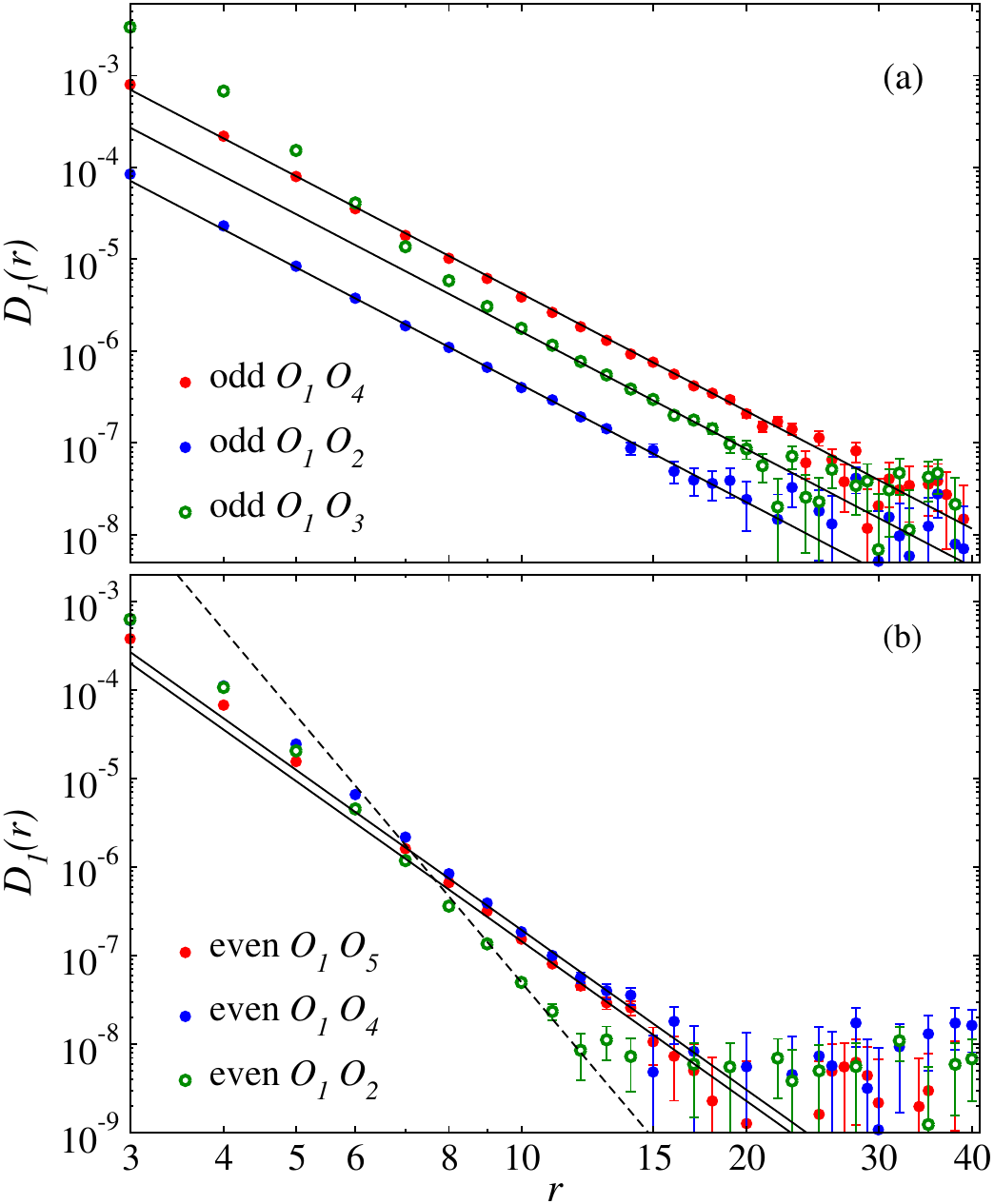}
\caption{Dependence on the choice of operators, illustrated by the second eigenvalues $D_{1}$ obtained when diagonalizing $2\times 2$ covariance matrices
in the odd (a) and even (b) $Z_2$ sectors. In each case, the operators used are indicated in the legends in the notation of Fig.~\ref{ops_is}. The
solid lines show the $\propto r^{-2\Delta_1}$ behaviors expected asymptotically for the first descendant operators, with $\Delta_1=17/8$ in (a) and
$\Delta_1=3$ in (b). The dashed line in (b) shows the behavior expected for the second even descendant, which has $\Delta_2=5$.}
\label{cor2c1}
\end{figure}

The choise of operators for given $N$ can also play a crucial role in optimally resolving the scaling dimensions. The six operators used here in both $Z_2$
sectors of course do not exhaust all possibilities on the $3\times 3$ cells but represent intuitive choices based on small numbers of $\sigma_i$ operators.
When using $N$ smaller than the full number of operators implemented in the simulations, different combinations of operators can be investigated post simulation.
Fig.~\ref{cor2c1} shows examples using $N=2$ with three different even or odd operator pairs from those in Fig.~\ref{ops_is}. Only the second eigenvalue is graphed
in each case, corresponding to the first descendant dimension $\Delta_1$. In the case of the odd sector in Fig.~\ref{cor2c1}(a) the correct decay 
exponent is well reproduced in all cases, but with the combination $(O_1,O_3)$ the corrections for small $r$ are much more prominent. The best choice among
the three pairs is $(O_1,O_4)$, which produces the largest overall magnitude of the eigenvalues and, therefore, the smallest relative statistical errors.
The scaling corrections for small $r$ are  also small.

In the case of the even operators in Fig.~\ref{cor2c1}(b), the pair $(O_1,O_2)$ does not produce
the expected decay governed by $\Delta_1$, but for the larger $r$ values, before the statistical cut off, the decay exponent is instead close to that
expected with $\Delta_2$. It is likely that the overlap with the first CFT descendant is very small with these two operators, while
being relatively large with the second descendant. The cross-over to the asymptotic form governed by $\Delta_1$ will then take place only for larger distances.
This potential problem should not be serious for larger $N$, where better overlaps with the most important operators should be expected. The variations
of the results with $N$ can also be regarded as an opportunity to optimize the performance of the method.

These examples show that the set of operators included in the covariance matrix can have large effects on the quality of the results. Since
the size $N$ of the full set used in a simulation has to be rather small, because of the $N^2$ scaling, the operators included should 
be considered carefully by testing, before large scale simulations are carried out.

\subsection{Small system sizes $L$ and $r = L/2$}
\label{sec:issmall}

In $d=2$ classical systems it should in general not be difficult to realize the range of distances $1 \ll r \ll L$ for which pure power law decays can be
observed. However, in higher dimensions, especially in the context of electronic models in $d=2+1$ dimensions simulated with auxiliary field quantum MC
methods \cite{assaad22,liu21,liu24,demidio24}, only relatively small systems can be accessed in practice. In conventional studies of correlation
functions it is then useful to set $r$ to a fraction of the system length $L$, e.g., to study correlations versus $r=L/2$. Though the correlations there
are maximally enhanced by the periodic boundary effects, the decay power is still the same as for $r \ll L$ in a large system. The utility of the covariance
method will of course be much broader if it can also be used without modifications for scaling with $r \propto L$. Since the
$L\times L$ square lattice does not have the ideal conformal geometry, the CFT arguments for disentangling of scaling dimensions in Sec.~\ref{sec:cftcov}
may not a priori hold unless $r \ll L$. However, one would still expect the same scaling dimensions to be at play, for the same reason that the standard
correlation functions decay with the correct critical exponents $2\Delta_0$ at $r \propto L$. There is also nothing specific about the arguments
in Sec.~\ref{sec:cftcov} that would point to them not remaining valid for all covariance eigenvalues for $r \propto L$, as long as $L$ is
sufficiently large and the decay of all correlations of CFT operators still are of the form Eq.~(\ref{cij}). We thus proceed to study the
covariance eigenvalues of small Ising lattices at $r=L/2$.

\begin{figure}[t]
\includegraphics[width=81mm]{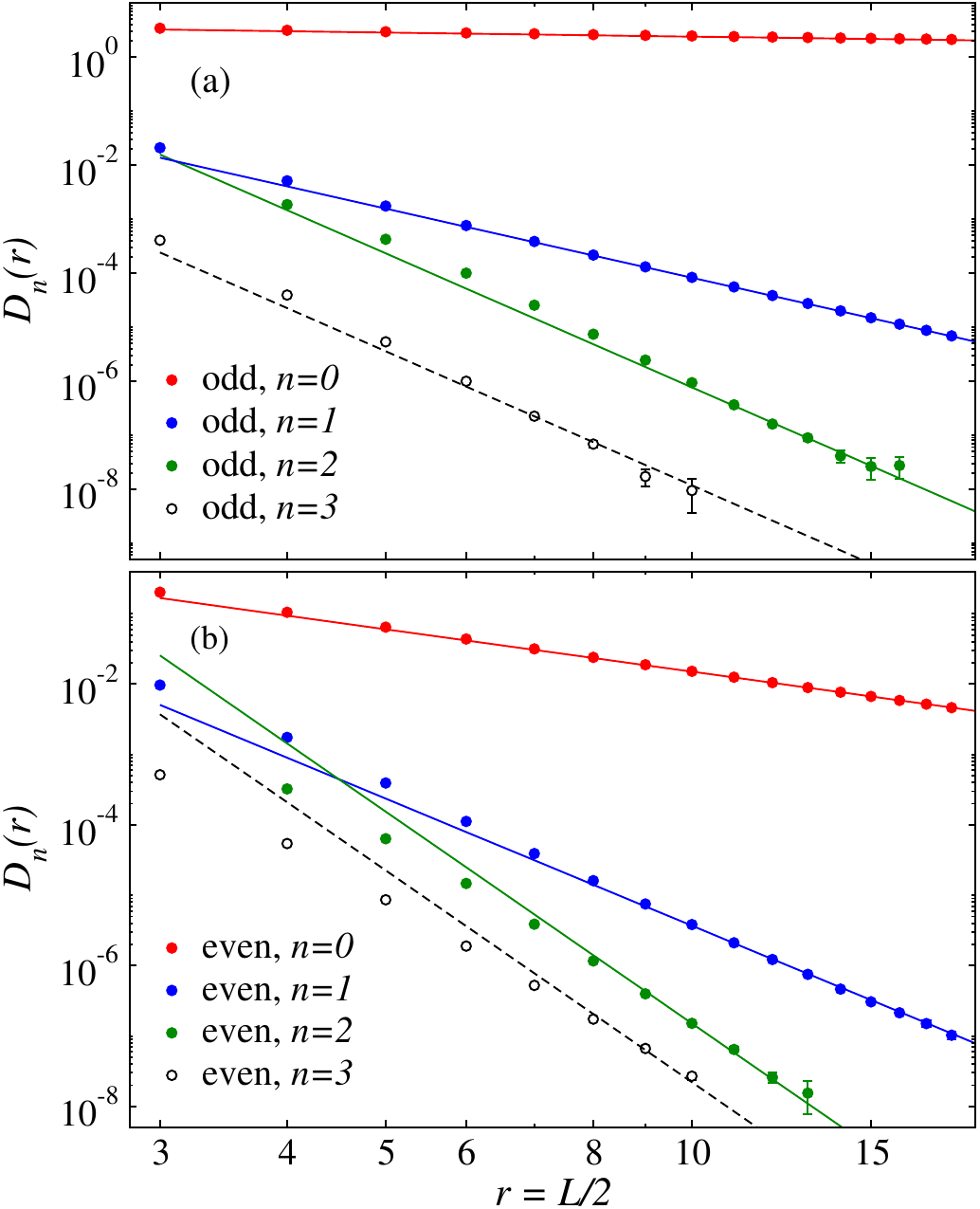}
\caption{Eigenvalues of the $Z_2$ odd (a) and even (b) covariance matrices computed with all the operators in Fig.~\ref{ops_is}
at distance $r=L/2$ for system sizes from $L=6$ to $L=36$. The red, blue, and green lines drawn adjacent to the $n=0,1,2$ data have
the slopes expected with the scaling dimensions of the respective Ising primaries and their first two descendants, as in Fig.~\ref{n246}.
The dashed black lines adjacent to the $n=3$ data also have slopes corresponding to $\Delta_2$}.
\label{cor6L}
\end{figure}

\begin{figure}[t]
\includegraphics[width=80mm]{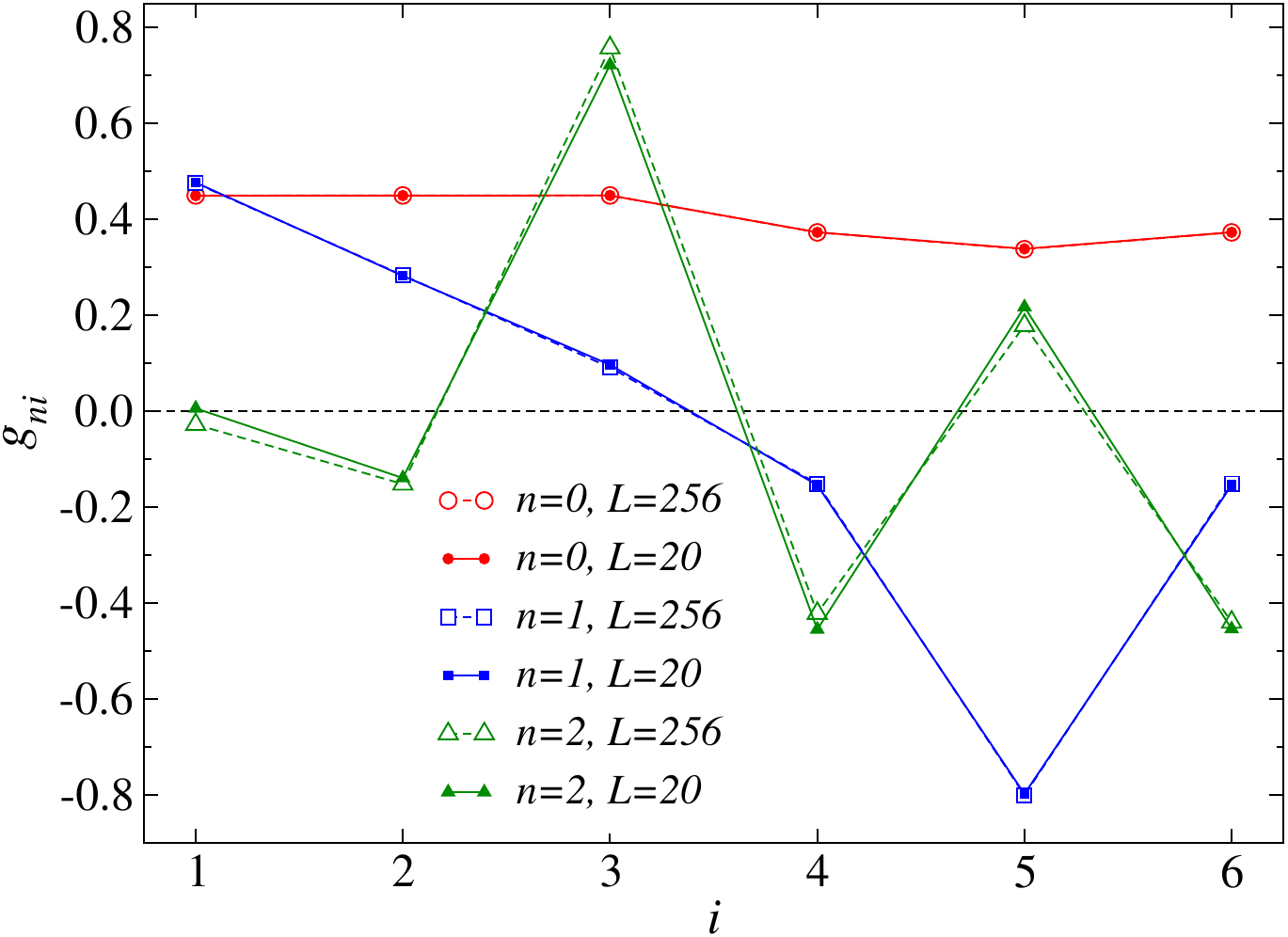}
\caption{Eigenvetors in the $Z_2$-odd sector corresponding to the scaling dimensions $\Delta_0$, $\Delta_1$, and $\Delta_2$, obtained from $r=10$ covariance
matrices with $N=6$ in a large system, $L=256 \gg r$ (open symbols), and in a small system of size $L=20$, where $r=L/2$. The corresponding eigenvalues
for $L=256$ and $L=20$ are shown in Figs.~\ref{n246}(e) and Fig.~\ref{cor6L}(a), respectively. The two data sets for $n=0$ and $n=1$ overlap to a large extent
while the minor differences between the $n=2$ sets reflect the statistical errors for $L=256$.}
\label{vec12}
\end{figure}

Figure \ref{cor6L} shows results for system sizes up to $L=36$ versus $r=L/2$, using covariance matrices with all the operators in Fig.~\ref{ops_is}. The
results do look similar to those in Fig.~\ref{n246}, but for these small systems also the fourth ($n=3$) eigenvalue can be well resolved beyond statistical
errors because the overall larger amplitudes of the boundary enhanced eigenvalues. The decay exponent is not the expected $2\Delta_3$, however, but instead is
very close to $2\Delta_2$ for the largest several system sizes in both $Z_2$ sectors. This failure to capture $\Delta_3$ likely just reflects transient behavior,
and crossovers to the expected forms should take place for larger systems. The eigenvectors for sufficiently large $r$ also are the same in systems with
$r \ll L$ and $r=L/2$, as shown in Fig.~\ref{vec12} for the case of $n=0,1$, and $2$ in the $Z_2$-odd sector. All these results indicate that the disentangling
of the scaling dimensions works independently of the system geometry.

\section{3D Ising model} \label{sec:is3}

\begin{figure}[t]
\includegraphics[width=60mm]{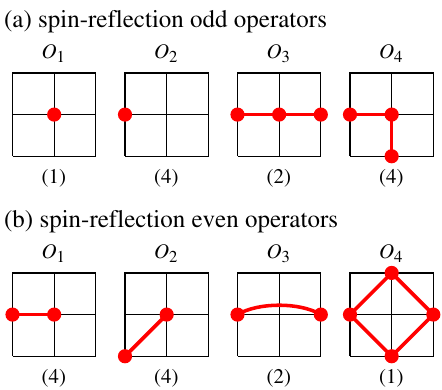}
\caption{$Z_2$ antisymmetric (a) and symmetric (b) operators on $3\times 3$ plaquettes used for the covariance matrices of the 3D Ising
model. The red circles and lines correspond, respectively, to spin operators and products as in Fig.~\ref{ops_is}, and the number of cell rotations
used to construct point-group symmetric operators is indicated beneath each cell.}
\label{ops_3d}
\end{figure}

\begin{figure}[t]
\includegraphics[width=45mm]{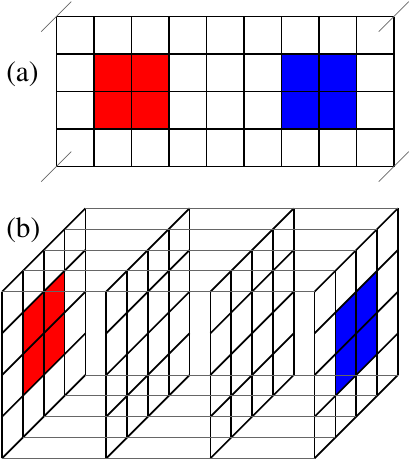}
\caption{Schematic illustration of correlation functions of the 3D Ising model defined with the 2D $3\times 3$ site operators in
Fig.~\ref{sep_3d}. In (a) both operators reside in the same $xy$ plane (corresponding to space-like separations in a corresponding quantum system)
at separation ${\bf r}=(5,0,0)$, while in (b) the operators are in different planes $z$ (time-like separation) but with the same in-plane $xy$ coordinates,
with ${\bf r}=(0,0,3)$.}
\label{sep_3d}
\end{figure}

The scaling dimensions of the 3D Ising model are not known rigorously, but recent works with the numerical conformal bootstrap method have produced
high-precision results that are believed to be exact within small windows of uncertainties originating from the ``islands'' of allowed values \cite{poland19}.
Here we use the same MC methods as in Sec.~\ref{sec:is} to study the Ising model on the simple cubic lattice with $L^3$ spins at its critical point,
testing scaling dimensions in both even and odd $Z_2$ sectors using the operators depicted in Fig.~\ref{ops_3d}. These purely in-plane operators are similar
to the ones studied above in the case of the 2D Ising model, some of them being identical to those in Fig.~\ref{ops_is}. While we could also study fully
symmetric 3D operators, we here take the opportunity to connect to quantum models in $d=2+1$ dimensions, where it is normally not practical to define
operators with a finite extent in imaginary time. Even if such definitions could be implemented, there is still not complete space-time symmetry of
the model---only emergent Lorentz symmetry with a model dependent velocity relating space and time distances. With the in-plane operators of the classical
3D Ising model, we study separations that are, in the classical-quantum mapping, either purely space-like, as illustrated in Fig.~\ref{sep_3d}(a),
or purely time-like, as in Fig.~\ref{sep_3d}(b), when considering the $z$ direction as discretized time in a quantum model.

The lack of full 3D symmetry of the operators implies that they do not correspond to a fixed Lorentz spin $l$ but mix $l=0$ and $l=2$ contributions.
The relevant operators are in the $l=0$ sector, with the symmetric ($\epsilon$) and antisymmetric ($\sigma$) ones  having scaling dimension
$\Delta_\epsilon = 1.412625(10)$ and $\Delta_\sigma = 0.5181489(10)$, respectively, where the numbers within $()$ indicate the uncertainty in the preceding
digit \cite{poland19}. The following primaries in these sectors have dimensions $\Delta_{\epsilon'} = 3.82968(23)$ and $\Delta_{\sigma'} = 5.2906(11)$.
Moreover, in the $l=2$ sector the energy-momentum tensor is marginal, i.e., with scaling dimension $\Delta_T=3$, and in the same sector the dimension
of following operator is $\Delta_{T'}=5.50915(44)$. Since descendants of all these operators exist in both the $l=0$ and $l=2$ sectors, 
pairs of eigenvalues with the same scaling dimension $\Delta_X + 2n$ should be expected for all the primaries $X$. With only four operators in the
covariance matrices, we can at most detect the four lowest scaling dimensions in each of the two $Z_2$ sectors. We only study relatively
small systems at $r=L/2$ in this case.

\begin{figure*}[t]
\includegraphics[height=98mm]{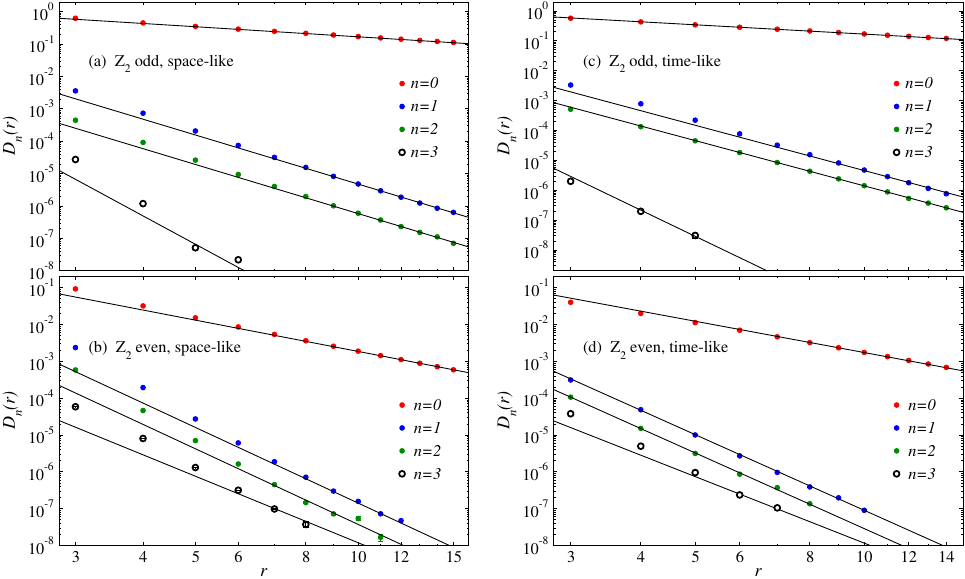}
\caption{Covariance eigenvalues of the 3D Ising model vs $r=L/2$ an $L \le 30$.
The operators included are defined in Fig~\ref{ops_3d}. Results from space- and time-like
correlations are shown in the graphs to the left (a,b) and right (c,d), respectively, corresponding to relative orientation of the operators according to the
illustrations in Fig.~\ref{sep_3d}. The lines fitted to the $n=1$ data sets in (a) and (c) show the expected scaling with the known value of
$\Delta_\sigma \approx 0.518149$, while those for the $n=2,3$ sets correspond to the $l=0$ and $l=2$ descendants, both with scaling dimension $\Delta_\sigma+4$.
Similarly, the solid lines in (b) and (d) show the expected scaling with the known value of $\Delta_\epsilon \approx 1.41262$ and the corresponding decsendant
dmensions $\Delta_\epsilon+4$. The dashed line in (d) corresponds to the expected slope with the scaling dimension $\Delta_T=3$ of the energy-momentum tensor.}
\label{cor3d}
\end{figure*}

The critical temperature of the 3D Ising model is known to high precision from large-scale MC simulations; $T_c = 4.51152325(10)$ \cite{ferrenberg18}.
Figs.~\ref{cor3d}(a,b) and \ref{cor3d}(c,d) show space-like and time-like eigenvalues, respectively, at criticality. The results overall are very similar
and the primaries and doubled first descendants can be convincingly detected in the decays for the largest distances $r=L/2$. However, the time-like separation
appears to have an advantage in smaller scaling corrections, which intuitively may be related to the fact that the time-like operator pairs are more
symmetric, maintaining the symmetries of the individual operators. Moreover, all the site-site distances between the two operators are closer to the
center-center distance $r$ on the case of time-like separations. For these reasons, most likely, it will also be better to study purely time-like
correlations in quantum systems.

In the case of the 2D Ising model, though the $n=2$ and $n=3$ data sets in Fig.~\ref{cor6L} also appear to be governed by the same scaling dimension, this
is most likely a non-asymptotic feature of the $n=3$ set, which is rather noisy, and the $n=4$ set (not shown) is completely noise dominated. In contrast, in
the case of the 3D Ising model, in Fig.~\ref{cor3d} the $n=1,2$ data sets develop either the same decay after the visible corrections have decayed away or
display no significant corrections. Moreover, the $n=3$ sets in Figs.~\ref{cor6L}(a) and \ref{cor6L}(c) clearly decay much faster, close to what is expected
with $\Delta_3$. There is, thus, little doubt that the $n=1,2$ pairs really reflect the first $l=0$ and $l=2$ descendants with common scaling dimensions.

In the $Z_2$ symmetric sector, the scaling dimension $\Delta_T$ of the energy-momentum tensor is smaller than the first descendant dimensions
$\Delta'_\epsilon$, and in Fig.~\ref{cor3d}(d) there is at least some indication of the fourth eigenvalue decaying in the way expected with $\Delta_T$.
In in Fig.~\ref{cor3d}(c), the decay of the fourth eigenvalue is much faster, close to what is expected with $\Delta_\sigma''$. In the case of the
space-like eigenvalues, for $n=3$ in Fig.~\ref{cor3d}(b) the decay in the available range of $r$ is somewhat faster than the expected $r^{-2\Delta_T}$ form,
being close to the same $r^{-2\Delta'_\epsilon}$ as the $n=1,2$ pair. This behavior most likely again just reflects slower convergence to the asymptotic
form than the corresponding time-like eigenvalue. The fourth $Z_2$-even eigenvalue decays rapidly but the behavior expected with $\Delta''_\sigma$ cannot
be fitted to the small number of points available.

Despite the large uncertainty in the fourth eigenvalues, the $n=0,1,2$ data demonstrate that the covariance method also works with $r=L/2$ in 3D lattices.
The better results from time-like correlations point to likely successful applications also to quantum criticality in $d=2+1$ dimensions.

\section{Blume-Capel model}
\label{sec:bc}

At a multiciritical point, two or more model parameters have to be tuned in order for all the relevant perturbations (scaling fields) to vanish. Lattice
operators will in general contain all the relevant scaling fields. Separating them reliably by fitting correlation functions to multiple power laws is in
difficult and unreliable, considering also the presence of scaling corrections. While corrections are also present in the covariance matrix, as we saw in
the previous sections, the asymptotic behaviors of each eigenvalue reflects a different scaling dimension. This way of disentangling scaling
dimensions may help immensely in studies of multicriticality, where two or more operators with the same symmetries as the Hamiltonian are relevant.
Here this capability of the covariance method will be exemplified by a study of the tricritical Ising universality class with the 2D BC model defined in
Eq.~(\ref{bcham}). The Ising coupling has been set at unity and the two tuning parameters are the temperature $T$ and the chemical potential $\lambda$.

The model parameters $(\lambda_t,T_t)=(1.965815,0.608578)$ from transfer-matrix calculations \cite{deng05} was used for the tricriticality
results presented here, but nearby points were also studied, including the estimate $(1.9660,0.6080)$ of the tricritical point obtained from Wang-Landau
MC simulations \cite{kwak15}. While there are some minor differences in the results, the decay exponents and eigenvectors are stable for the system sizes
considered here---further results supporting this stability of the scaling dimensions  will be presented in Sec.~\ref{crossover} in the context of the
crossover from tricritical to regular Ising criticality on the continuous part of the phase boundary away from $(\lambda_t,T_t)$.

\begin{figure}[t!]
\includegraphics[width=75mm]{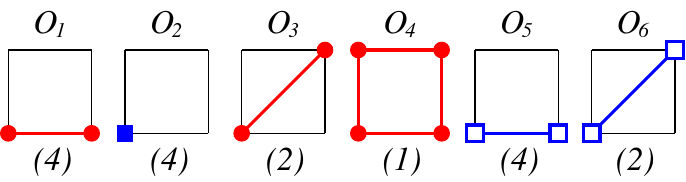}
\caption{Operators on $2\times 2$ plaquettes used for the covariance matrices of the BC model. Red circles represent spin operators taking the values
$\sigma_i \in \{-1,0,+1\}$, blue solid squares $\sigma_i^2 \in \{0,1\}$, open blue squares $1-\sigma_i^2$. The colored lines imply products
of the operators. The number of operations (cell rotations producing different patterns) used for symmetrizing is indicated beneath each plaquette.}
\label{ops_bc}
\end{figure}

Standard MC simulations on $L\times L$ periodic lattices were again carried out, in this case combining Swendsen-Wang cluster updates \cite{swendsen87} for
the $|\sigma_i| = 1$ spins with Metropolis updates to change the number $N$ of spins $|\sigma_i|=1$ and to interchange $\sigma_i=0$ vacancies and $|\sigma_j|=1$
spins at different locations $i,j$. An MC step entails the construction of all Swendsen-Wang clusters and flipping each of them with probability $1/2$, followed by
$N$ and $2N$ attempts, respectively, of single-site and two-site Metropolis updates. Six operators are again used for the covariance matrices, but here we
only consider the $Z_2$-even sector and use plaquettes of $2\times 2$ sites. The operators are illustrated in Fig.~\ref{ops_bc}.

Given that the autocorrelation time is quite long in this model (because the vacancies are affected only by the local Metropolis updates) and the $N^2$ scaling
of the covariance calculations, the matrices were accumulated only after every 100 MC steps in the case of the $L=256$ system of main focus (and more often for
smaller systems). The covariance matrices were again averaged over ${\bf r}=(r,0)$ and $(0,r)$. A bin average of the covariance matrix involves $5\times 10^6$
MC steps in this case and about $10^4$ bins were generated.

The tricritical Ising point corresponds to $m=4$ in the CFT hierarcy in Eq.~(\ref{cftc1}).
The values of the primary scaling dimensions corresponding to the fully symmetric operators
studied here are, in the notation of  Eq.~(\ref{cftc1}), $\Delta_{33} =1/5$, $\Delta_{32}=6/5$, and $\Delta_{31}=3$ (or $2h_{\epsilon}$, $2h_{\epsilon'}$, and
$2h_{\epsilon''}$ with the conventions of Ref.~\onlinecite{yellowbook}). Out of these, $\Delta_{33}$ and $\Delta_{32}$ are relevant and tuning them both to zero
in a microscopic model will in general require two adjustable parameters; $\lambda$ and $T$ in the BC model. 

\begin{subequations}
Since $\Delta_{33}$ and $\Delta_{32}$ differ
exactly by $1$, one should expect contributions to correlation functions that decay as $\propto r^{-2\Delta}$ with $\Delta$ taking all the values $1/5 + a$ for
$a=0,1,2,\ldots$, as well as $3 + b$ with $b=0,2,\ldots$. For simplicity of the notation, we will here just use $\Delta_i$, $i=1,2,\ldots$ to refer to all
the scaling dimensions in order of increasing values (instead of $n=0,1,\ldots$ used above for the Ising models, which has only one primary, $n=0$, in each $Z_2$
sector), and those that we in principle should be able extract using $6\times 6$ covariance matrices are
\begin{eqnarray}  
  \Delta_1 & \equiv \Delta_{33} = 0.2, \\
  \Delta_2 & \equiv \Delta_{32} = 1.2, \\
  \Delta_3 & \equiv \Delta'_{33} = 2.2, \\
  \Delta_4 & \equiv \Delta_{31} = 3.0, \\
  \Delta_5 & \equiv \Delta'_{32} = 3.2, \\
  \Delta_6 & \equiv \Delta''_{33} = 4.2,
\end{eqnarray}  
\label{deltavals}
where primes refer to descendants. Similarly, we label the eigenvalues for given $r$ by $D_i(r)$ with $i=1,\ldots,N$ in increasing order. For large $r$, the
indices for the scaling dimensions and the eigenvalues should then match; $D_i \sim r^{-2\Delta_i}$. For small $r$, the ordering of the eigenvalues may again
be different in some cases, as in Fig.~\ref{n246}. We will see some evidence of such crossing behavior of two of the eigenvalues for $r$ beyond the largest
accessible distance.
\end{subequations}

\subsection{$N=2$ eigenvalues and the tangent vector of the phase boundary}
\label{sec:N2}

\begin{figure}[t!]
\includegraphics[width=80mm]{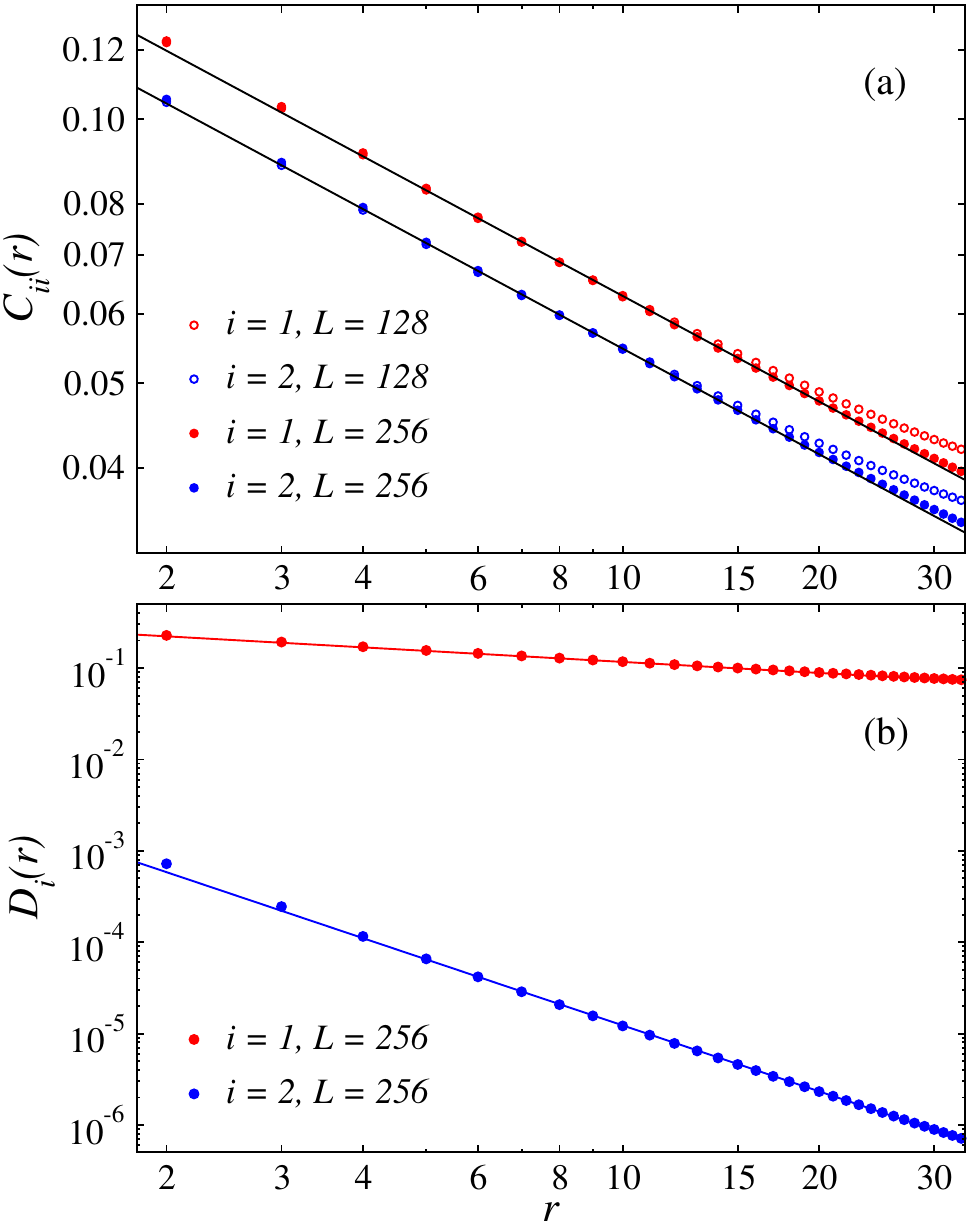}
\caption{(a) Distance dependent correlation functions of the individual operators $O_1$ and $O_2$ in Fig.~\ref{ops_bc} computed on $L=128$ and $256$
lattices at the estimated tricritical point of the BC model. The lines show the power-law decays expected with the scaling dimensions of the smallest primary
operator $\Delta_1=1/5$. (b) Eigenvalues of the distance dependent covariance matrix of $O_1$ and $O_2$ for $L=256$. The lines show the
expected power laws with the scaling dimensions $\Delta_1=1/5$ and $\Delta_2=6/5$ .}
\label{corbc2}
\end{figure}

We first consider the $2\times 2$ covariance matrix of $O_1$ and $O_2$ in Fig.~\ref{ops_bc}, which make up the $E$ and $F$ terms, respectively, in the Hamiltonian
Eq.~(\ref{bcham}). The diagonal elements (the conventional correlation functions) are graphed versus distance in Fig.~\ref{corbc2}(a). To test convergence
with the system size, results for $L=128$ and $256$ are compared. The periodic boundary conditions again cause enhancements of the correlations that
grow with $r$, but these boundary effects can largely be avoided at the shorter distances. There the two correlation functions are, as
expected, completely dominated by the contribution from the smallest scaling dimension $\Delta_1$, but some deviations from the leading power law are
visible at short distances. While there are finite-size effects left even in the limited range of $r$ in Fig.~\ref{corbc2}(a), for $L=256$ 
there is good agreement with the expected leading asymptotic exponent for $r \in [6,20]$.

Figure \ref{corbc2}(b) shows the two eigenvalues of the covariance matrix. The decay forms are indeed consistent with the two leading
scaling dimensions $\Delta_1$  and $\Delta_2$, though there are, as expected, corrections at the shortest distances. The eigenvectors $g_i=(g_{i1},g_{i2})$
defining $Q_i=g_{i1}O_1+g_{i2}O_2$ are $g_1=(g_{11},g_{12})=(0.731,0.682)$ and $g_2=(g_{21},g_{22})=(0.682,-0.731)$ in the range $r \in [5,20]$, with
very small changes for larger distances. 

The eigenvectors mixing the two energy terms in the Hamiltonian should give the special directions in the phase diagram along which critical scaling 
is governed only by the corresponding correlation length exponents $\nu_1=(2-\Delta_1)^{-1}$ and $\nu_2=(2-\Delta_2)^{-1}$. The above eigenvalues 
were, however, computed with relative weights different from those of $E$ and $F$ defined in Eq.~(\ref{bcham}). The proper weighting of the two
operators is obtained with $2O_1$ and $O_2$ (corresponding to $2L^2$ Ising bond operators and $L^2$ single-site operators on the lattice),
which gives the eigenvectors $g_1=(0.906,0.423)$ and $g_2=(0.423,-0.906)$ (i.e., the ratio of the 
coefficients trivially change by a factor of $2$). With the Ising interaction strength set to unity in Eq.~(\ref{bcham}), one of these vectors should 
define the tangential direction at the tricritical point of the phase boundary expressed as $\mu_c=-(\beta\lambda)_c$ versus $\beta$, where 
$\beta \equiv T^{-1}$. Examining the phase boundary $\lambda_c$ versus $T$ in Fig.~1 of Ref.~\onlinecite{kwak15} (which uses slightly different notation), a 
derivative $d\lambda_c/dT \approx -0.28$ at the tricritical point can be estimated. This translates to $d\mu_c/d\beta \approx -2.17$, which 
should equal the ratio $g_{i2}/g_{i1}$ for one of the eigenvectors of the covariance matrix. Indeed, the derivative closely matches
$g_2$ (corresponding to $\Delta_2=6/5$), with $g_{22}/g_{21} \approx -0.906/0.423 \approx -2.14$. This agreement within less than $2$\% is as good
as could possibly be expected, given the uncertainty of the tangent estimated from the phase diagram in Ref.~\onlinecite{kwak15}. The present result
for the tangent should be the best available so far, as it avoids the difficulties of constructing a challenging portion of the phase boundary
to high precision.

The fact that $\Delta_2$, not $\Delta_1$, is the relevant scaling dimension here is also expected on account of the fact that the larger of the
correlation length exponents, here $\nu_2=(2-\Delta_2)^{-1}$, in general governs the divergence of the correlation length on the line tangential
to the phase boundary at a tricritical point.

\subsection{$N=6$ eigenvalues for $r \ll L$}

\begin{figure}[t!]
\includegraphics[width=84mm]{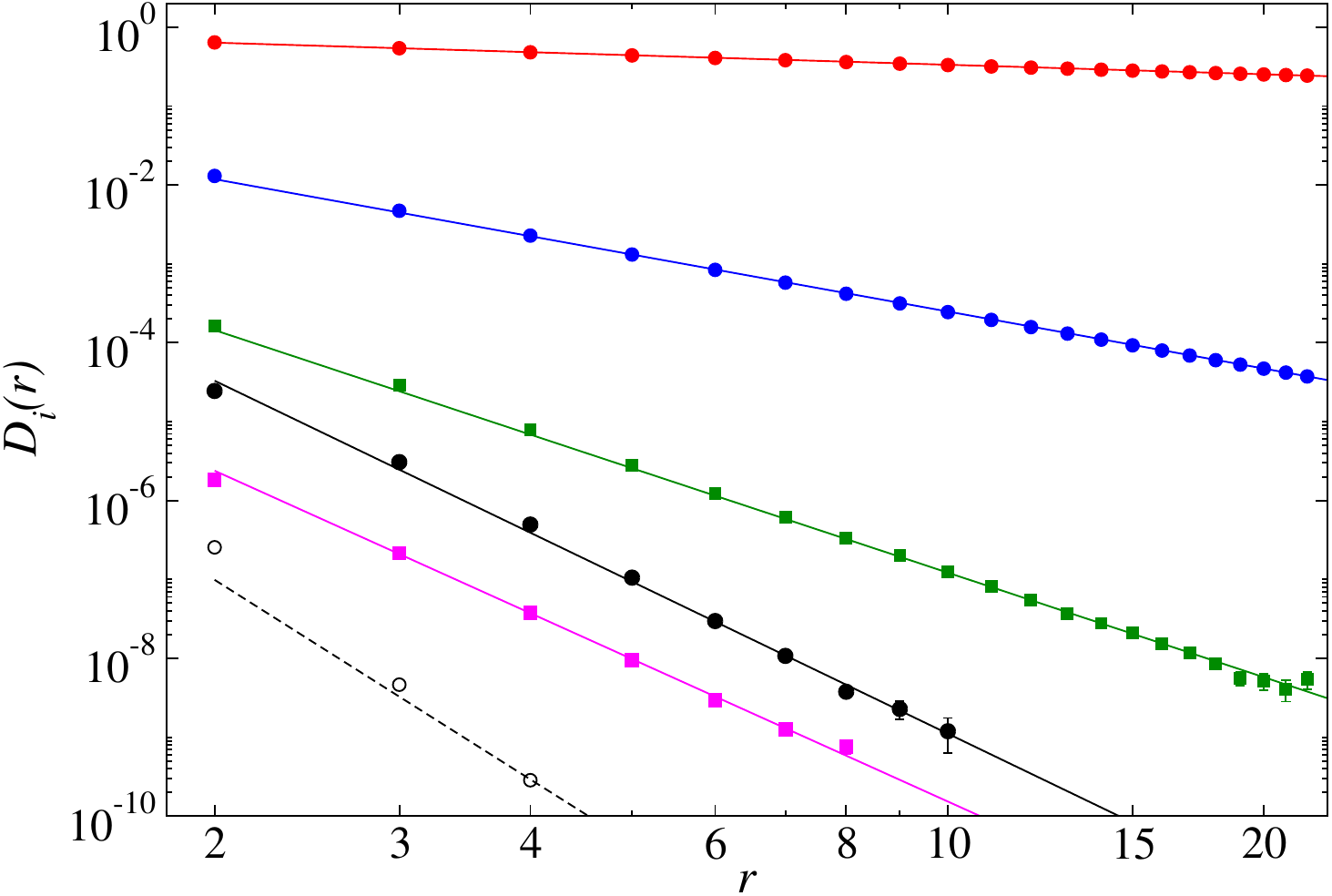}
\caption{(a) Eigenvalues vs distance of the BC covariance matrix of all the operators depicted in Fig.~\ref{ops_bc}. The fitted lines
show power laws with the exponents $2\Delta_i$ expected with the BC scaling dimensions listed in Eqs.~(\ref{deltavals}). The lines correspond
to $i=1,2,3,5,4,6$ from top to bottom.}
\label{corbc6}
\end{figure}

Moving on to the full $6\times 6$ covariance matrix, Fig.~\ref{corbc6} shows $L=256$ results for $D_i(r)$ up to $r=22$, where the boundary effects
are small. Remarkably, the six lowest exponents ($2\Delta_i=0.4$, $2.4$, $4.4$, $6$, $6.4$, and $8.4$) are all well reflected in the observed power-law
decays, though in the case of $i=6$ the decay is too fast to capture the asymptotic form, with large corrections likely affecting the three available data
points. The $i=4$ and $i=5$ data sets appear to correspond to the scaling dimensions $\Delta_5$ and $\Delta_4$, i.e.,
in the reverse order, as we saw examples of in the case of the Ising model in Sec.~\ref{sec:is}. However, in Fig.~\ref{corbc6} an eventual
crossing of the eigenvalues for much larger $r$ is less clear, given the closeness of the scaling dimensions $\Delta_4=3$ and $\Delta_5=3.2$,
and it is also possible that the convergence to the ultimate power laws is just slow and there is no crossing of the eigenvalues.

The $i=1,\ldots,5$ data sets can be fitted to power laws, which we do here in addition to showing the power laws with the known exponents $2\Delta_i$
in Fig.~\ref{corbc6}. We fit without corrections (additional power-law terms), including a range of $r$ values for which the corrections appear
to be relatively small and high quality fits can be obtained. A more careful procedure would consider corrections and also systematic
investigations of small remaining finite-size effects (which are important only for the slowest decaying modes). The fitting results here
should just be taken as examples of reasonable fits, not an exhaustive analysis.

Since the data for different $r$ values are correlated, a statistically sound fitting procedure must include the full covariance matrix of the MC
fluctuations, instead of just the conventional standard deviations used for uncorrelated data. Note that such an ``error covariance matrix''
$C_{\rm err}(r,r')$ involves a range of distances included in the fits, $r,r'  \in [r_{\rm min},r_{\rm max}]$, for the same eigenvalue and is very different
from the operator covariance matrix $C_{ij}(r)$. Some of the eigenvalues of $C_{\rm err}(r_i,r_j)$ are typically very small, i.e., there are modes with very
small fluctuations. Including all of them in the definition of the goodness-of-fit $\chi^2$ is not possible, because very small deviations of the true
functional form from a single power law (and such deviations cannot be eliminated when fitting without any corrections to scaling) can be reflected in
very large $\chi^2$ values. A meaningful fit to a single power law then must include only a few (some times only two) $C_{\rm err}(r_i,r_j)$ eigenmodes.
The final error bars on the exponents can be estimated by repeating the fits multiple times with Gaussian noise added to the eigenvalues of
$C_{\rm err}(r_i,r_j)$.

The fitting windows $r \in [r_{\rm min},r_{\rm max}]$ are here chosen with $r_{\rm min}$ sufficiently large to eliminate most of the scaling corrections
and with $r_{\rm max}$ small enough to avoid large boundary effects and to exclude data with error bars too large to be useful. For some consistency
among the different data sets, the fits reported below are for $r \in [8,16]$ for data sets $i=1,2,3$, while for sets $i=4,5$, where only a small
number of data points are available, $r \in [6,10]$ for $i=4$ and $r \in [5,9]$ for $i=5$; for the latter two sets $r_{\rm max}$ is the largest $r$
shown Fig.~\ref{corbc6}. The results for the scaling dimensions are then, with statistical errors representing one standard deviation:
$\Delta_1 = 0.1993 \pm 0.0002$, $\Delta_2 = 1.1973 \pm 0.0002$, $\Delta_3 = 2.22 \pm 0.02$, $\Delta_4 = 3.24 \pm 0.15$, and $\Delta_5 = 2.98 \pm 0.12$.

Comparing with the expected values in Eq.~(\ref{deltavals}), the two leading scaling dimensions are very close to their exact values, with deviations of only
about $0.4\%$ and $0.3\%$, respectively, for $\Delta_1$ and $\Delta_2$. In the case of $\Delta_2$, the deviation represents more than 10 standard deviations,
however, reflecting some remaining systematical errors from scaling corrections, as discussed in Sec.~\ref{sec:lattcov}. Changing the fitting window
slightly does not affect the result appreciably. The result for $\Delta_3$ is within $1\%$ of the correct value, in both absolute and statistical
terms. The $i=4,5$ results have large error bars but do not otherwise deviate from their expected values, apart from the fact that the order of these
eigenvalues is switched in the range of $r$ for which reasonably good data are available. It is again not possible to conclude with certainty that the
order of these eigenvalues really is reversed.

It would be impossible to extract all of these scaling dimensions, to the level of statistical precision achieved here, from a conventional correlation
function. At most the two smallest dimensions can be extracted by fitting power laws, at worse fidelity than achieved here. While there is some uncertainty in
the data corresponding to $\Delta_4$ and $\Delta_5$, and the $i=6$ data set only shows rough consistency with $\Delta_6$, the good results for $\Delta_3$
is particularly noteworthy.

The eigenvectors corresponding to the three smallest scaling dimensions are
\begin{equation}
g_1 \approx \left ( \begin{array}{r} 0.436 \\ 0.407 \\  0.444 \\ 0.403 \\ -0.379 \\ -0.375 \end{array} \right )~ 
g_2 \approx \left ( \begin{array}{r} 0.179 \\ -0.138 \\  0.328 \\  0.549 \\ 0.448 \\ 0.583  \end{array} \right )~
g_3 \approx \left ( \begin{array}{r} 0.262 \\ 0.017 \\  0.616 \\ -0.715 \\  0.139  \\  0.144 \end{array} \right ), \nonumber
\end{equation}
where the elements from top to bottom are the coefficients of $O_1,\ldots,O_6$ in Fig.~\ref{ops_bc}(b). Here $g_1$ and $g_2$ change 
very little with $r$, while $g_3$ is statistically noisy for $r>10$ but does not change appreciably between $r=4$ and $r=10$.
The above results are all for $r=10$ and have statistical errors beyond the digits shown in the case of $g_1$ and $g_2$, while
the $g_3$ elements have statistical errors of roughly $0.003$ (i.e., affecting the last digit displayed).

\begin{figure}[t]
\includegraphics[width=84mm]{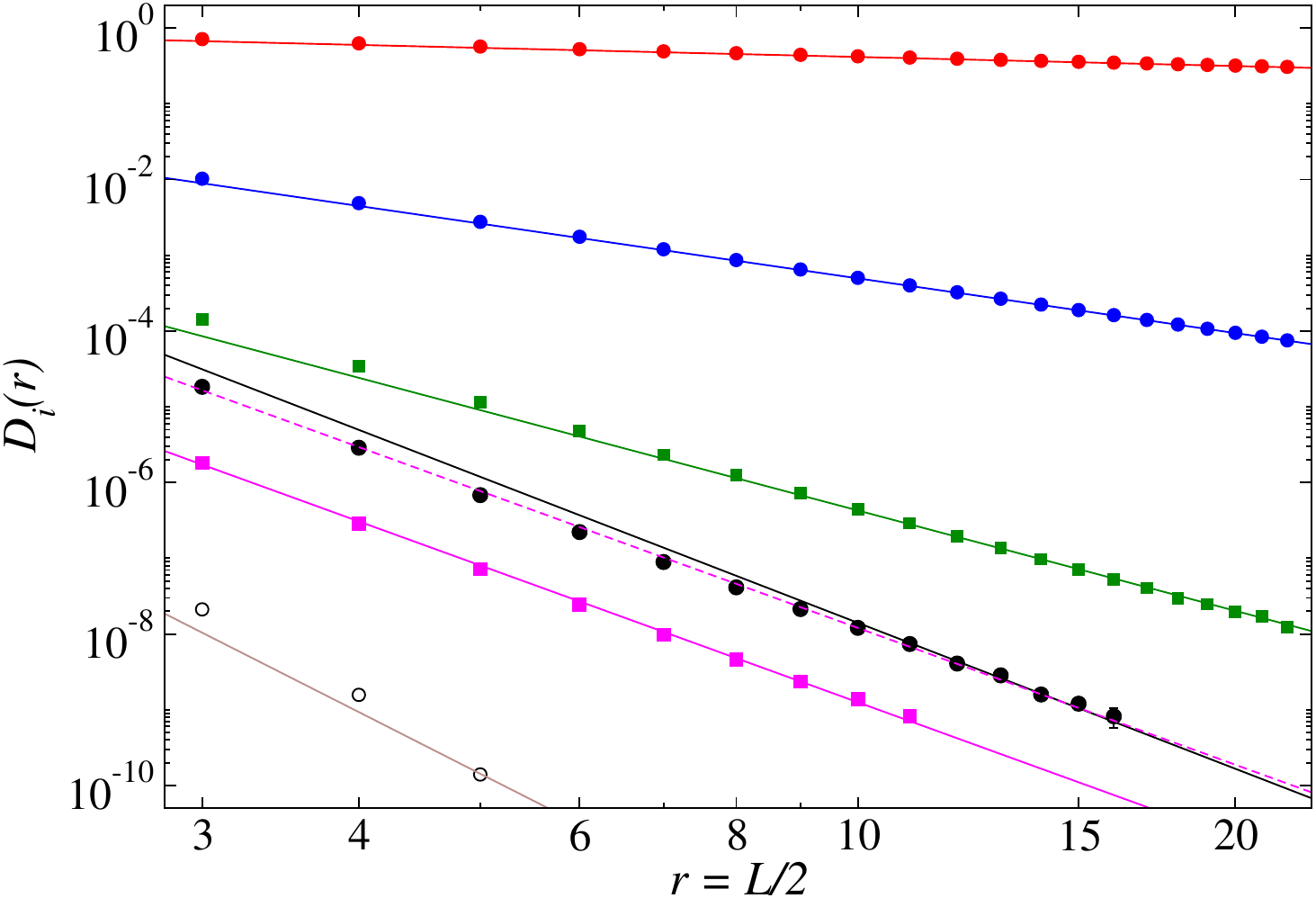}
\caption{Size dependent eigenvalues of the BC model vs $r=L/2$ for even lengths $L\le 44$. The solid lines show power laws with the exponents $2\Delta_i$
expected with the BC scaling dimensions listed in Eqs.~(\ref{deltavals}), with the index $i$ in the order $i=1,2,3,5,4,6$ from top to bottom.
The dashed magenta line shows a fit assuming that the $i=4$ and $i=5$ eigenvalues are not in the reverse order, i.e., with $\Delta_4$ governing
the $i=4$ set. If that is the case, the $i=5$ set must not yet be in its asymptotic regime governed by $\Delta_5$ (for which no line is shown
with the $i=5$ set).}
\label{corbcL6}
\end{figure}

\begin{figure*}[t]
\includegraphics[height=94mm]{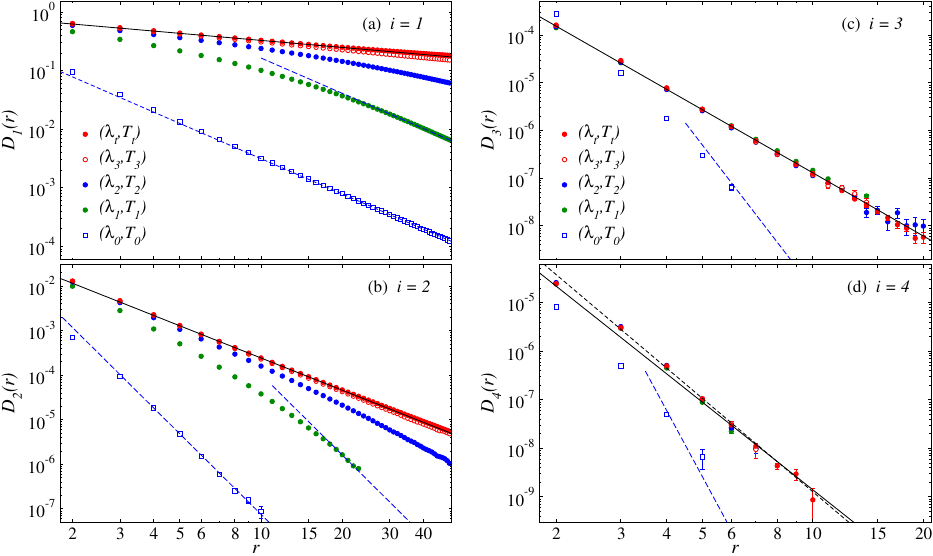}
\caption{The four largest eigenvalues of the $6\times 6$ covariance matrix of all the operators in Fig.~\ref{ops_bc} graphed vs the distance $r$
at five points $(\lambda_k,T_k)$ on (within the precision of the calculations) the part of the phase boundary of the BC model where the transitions are
continuous. In all the graphs, these points are $(\lambda_0,T_0)=(0,1.69356)$, $(\lambda_1,T_1)=(1.95,0.65778)$, $(\lambda_2,T_2)=(1.96253,0.62)$,
$(\lambda_3,T_3)=(1.96541,0.61)$, and $(\lambda_t,T_t)=(1.965815,0.608578)$ is the estimated tricritical point. The transition is of regular Ising type for
$k=0,\ldots,3$. The black lines have slopes corresponding to the expected tricritical exponents $2\Delta_i$ with $i=1,\ldots,4$ in (a),\ldots,(d), i.e.,
$2\Delta_i=0.4$, $2.4$, $4.4$, and $6$, respectively. In (d), the black dashed line show the slope if the data set instead corresponds to $2\Delta_5=6.4$
(as the black line with the $i=4$ data set in Fig.~\ref{corbc6}). The blue dashed lines show the slopes corresponding the four smallest $Z_2$-even
Ising scaling dimensions, i.e., the exponents $2,6,10,14$ in (a),\ldots,(d), respectively. In (c) and (d), many of the plotting symbols are not
visible because of the closeness of the data for all cases but $(\lambda_0,T_0)$.}
\label{cross}
\end{figure*}

Unlike the $N=2$ case discussed earlier, for $N=6$ the relationship between the eigenvectors and the special directions in the two-dimensional phase 
diagram is nontrivial, given that the original operators $O_i$ are not orthogonal. It is interesting to note, however, that the eigenvector $g_1$ has
almost exactly the same ratio of coefficients, $g_{11}/g_{12} \approx 1.07$ as in the $2\times 2$ case, which is not true for the second eigenvector,
where now $g_{22}/g_{21} \approx -0.78$ (instead of $-1.07$ in the $2\times 2$ case).

\subsection{$N=6$ eigenvalues for $r = L/2$}

Eigenvalues for system sizes $L$ up to $44$ are shown in Fig.~\ref{corbcL6} versus $r=L/2$. The behaviors are very similar to those for $r \ll L$
(Fig.~\ref{corbc6}), as also found in the case of the 2D Ising model in Sec.~\ref{sec:is}. Here the $i=4$ and $i=5$ eigenvalues are not as clearly
associated with the scaling dimensions $\Delta_5$ and $\Delta_4$ in the reverse order, with the $i=4$ set also being well described by $\Delta_4$.
In Fig.~\ref{corbc6} the reversal of the eigenvalues is more convincing (though not completely clear) and suggests a crossing of the two power laws
for $r$ much larger than the maximum accessible distance. The $i=4$ and $i=5$ eigenvectors from the $r=L/2$ and $r \ll L$ calculations are very similar,
suggesting that the eventual crossing of these eigenvalues should either happen in both cases or in neither case. Given that the $i=5$ data set in
Fig.~\ref{corbcL6} matches $\Delta_4$ (which is the dimension of the third primary) quite well and the corrections to scaling even for the smallest
$r$ value are small, the most likely scenario is a slow convergence of the $i=4$ data set to the behavior governed by $\Delta_5$, which is the dimension
of the first descendant of the $\Delta_2$ operator. It is not clear, however, why the convergence of this eigenvalue should be slower in the $r=L/2$
case and also slower than that of the third primary dimension.

\section{Critical Crossover}
\label{crossover}

The tricritical Ising point $(\lambda_t,T_t)$ of the BC model studied in Sec.~\ref{sec:bc} is the end point of a curve $(\lambda_c,T_c)$ on which the
transition is of regular Ising type for $\lambda_c < \lambda_t$ and $T_c > T_t$. For $\lambda_c > \lambda_t$ and $T_c < T_t$ the transition is of first order.
On the critical part of the phase boundary, for $(\lambda_c,T_c)$ close to $(\lambda_t,T_t)$ the physical quantities should exhibit crossover behaviors, with
tricritical behavior applying up to a length scale $\Lambda \propto |\delta|^{-\nu_\parallel}$, where $\delta$ is the distance to $(\lambda_t,T_t)$ and
$\nu_\parallel$ is the exponent governing the correlation length when approaching the tricritical point in the tangent direction of the phase boundary. 
As discussed in Sec.~\ref{sec:N2}, this exponent is related to the larger of the two correlation-length exponents of the tricritical point,
$\nu_\parallel = (2-\Delta_2)^{-1} = 5/4$. On length scales larger than $\Lambda$, the regular Ising criticality should apply, and, therefore, a crossover
regime in$r$ also must exist where none of the critical behaviors is cleanly exhibited.

It is interesting to investigate how the crossover behavior is manifested in the eigenvalues of the covariance matrix. For this purpose, several values
of $(\lambda_c,T_c)$ on the critical phase boundary were determined, using the standard Binder cumulant method. Among these points, the one furthest away from
the tricritical point has zero chemical potential, $(\lambda_0,T_0)=(0,1.69356)$, where the transition temperature $T_0$ agrees well with a recent
result \cite{moueddene24}. The other points on the critical line are labeled $(\lambda_k,T_k)$, $k=1,2,3$, and are located much closer to the
tricritical point. The values of $\lambda_k$ and $T_k$ are listed in the caption of Fig.~\ref{cross}. This figure shows the four smallest
covariance eigenvalues (using the same six operators as in Sec.~\ref{sec:bc}) versus the separation $r$. Figs.~\ref{cross}(a)-\ref{cross}(d) correspond
to data sets $i=1,\ldots,4$ and the eigenvalues for all five critical points are graphed in the same panel, with lines showing the power laws corresponding
to both regular Ising and tricritical Ising behaviors.

In Figs.~\ref{cross}(a) and \ref{cross}(b), regular critical behavior is well manifested at the point $(\lambda_0,T_0)$, which is very far
from the tricritical point. The crossover to regular Ising behavior is also seen for $(\lambda_1,T_1)$ when $r$ exceeds $15 \sim 20$. However,
no clear-cut tricritical behavior is observed at smaller $r$; instead the flow to the ultimate regular Ising fixed point has already commenced at
the shortest distances. In contrast, at $(\lambda_3,T_3)$ the eigenvalues follow the tricritical behavior up to $r = 10 \sim 15$ and only
deviate slightly (though clearly visibly) from it up to the largest distances considered. At $(\lambda_2,T_2)$ the tricritical behavior is
less pronounced and the regular Ising criticality also has not set in for $r$ up to $40$. These behaviors of the first and second
eigenvalues are qualitatively expected.

The third and fourth eigenvalues in Figs.~\ref{cross}(c) and \ref{cross}(d) are more interesting. Surprisingly, here the tricritical behavior
is observed for all the points except for $(\lambda_0,T_0)$, and at the latter the results are noisy and the asymptotic regular Ising behavior is
not observed clearly. In Fig.~\ref{cross}(c) the last points are nvertheless consistent with the expected decay. In ~\ref{cross}(d) the decay with
$r$ is very fast but not yet as fast as the expected $r^{-14}$. The main new insight here is that the tricritical behavior of the third and fourth eigenvalues
is very stable for all the points $(\lambda_k,T_k)$, $k=1,2,3$, even though the regular Ising model should eventually set in, as it does already for
small $r$ in the case of $(\lambda_0,T_0)$ (which is not a special point even though $\lambda=0$). The stability of these eigenvalues is likely a
consequence of the fact that the corresponding regular Ising scaling dimensions are so large, thus the perturbations of the corresponding levels
in the tricritical CFT are very weak. 

\section{Conclusions and Discussion}
\label{sec:discuss}

The results obtained here demonstrate that the covariance method is capable of resolving multiple scaling dimensions in remarkable detail, far
beyond what is possible by fitting conventional correlation functions to a sum of power laws. In the conventional way of fitting individual
correlation functions, or even combining several correlation functions, the analysis is complicated by the fact that there are corrections of
the form $r^{-(\Delta_j-d)}$ from all irrelevant perturbations ($\Delta_j > d$), in addition to the contributions to the correlation functions from
several power laws $r^{-2\Delta_i}$. In practice, it is not possible to obtain reliable results for the fast decaying contributions from descendants
or higher primaries under these conditions. The covariance method solves this problem by disentangling the different $r^{-2\Delta_i}$ decays.
Thus, while the corrections from irrelevant perturbations are still present in the eigenvalues (and it may be useful to include corrections
when fitting), their asymptotic forms governed by different power laws makes it possible to extract multiple scaling dimensions.

A well known method for reducing scaling corrections is to introduce other interactions in a model so that the field corresponding to the leading
correction can be tuned away, or at least made very small \cite{ballesteros98,pelissetto02}. This method has helped to produce some of the best MC results
for critical exponents \cite{campostrini02,hasenbusch10}, but its original intent was not to extract irrelevant scaling dimensions or multiple relevant
dimensions. In cases where the descendants are of interest, tuning away a scaling correction may also be helpful, though fitting of the eigenvalues can also
be done with corrections. It should also be kept in mind that identifying the optimal interactions to minimize the scaling corrections can be
a very time consuming task.

The exact number of scaling dimensions that can be obtained reliably from the eigenvalues of course depends on the system under study. Here it was possible
to observe a scaling dimension as large $\Delta =5$ in the 2D Ising model and $\Delta \approx 3.5$ in the 3D ising model. In the latter case, the following
not yet resolved scaling dimension should be slightly larger than $5$ and may also be accessible with more (and more optimally selected) operators included
in the covariance matrix. At a multicritical point, here in the 2D BC model, the larger number of relevant primary operators, with their descendants,
implies a larger number of resolvable scaling dimensions with reasonably small values. To resolve many operators, a correspondingly sized covariance
matrices of course have to be used, and longer MC simulations may also be required to properly disentangling the many power laws with rather close
decay exponents.

The ability to extract multiple scaling dimensions and eigenfunctions corresponding to relevant orthogonal operators at a tricritical point
(and likely also at higher-order multicritical points) is also an appealing feature. There are many potential applications to multicriticality, e.g., further
numerical studies of deconfined quantum-critical points \cite{zhao20,chen24a,chester24,takahashi24} and liquid-gas transitions \cite{yarmo17,yarmo18}.
While only the 2D BC model was used here to study the tricritical Ising point and associated crossover behavior, this model is highly non-trivial and the
good results obtained for several scaling dimensions is highly suggestive of excellent performance more broadly.

An important feature of the covariance method is that it works not only for $r \ll L$ on large lattices, but also for $r \propto L$ in relatively
small systems. This aspect should be particularly useful in quantum MC studies of fermion systems, where the accessible system lengths are often only of the
order $20$ \cite{assaad22,liu21,liu24,demidio24}. In 3D classical models and 2+1 dimensional quantum spin models (those for which the negative sign
problem can be avoided \cite{henelius00}), larger system sizes can be studied. However, with small scaling dimensions the boundary effects can still be
problematic \cite{takahashi24} and studies for $r=L/2$ will be useful also in those cases. The method is easy to implement with any MC or quantum MC
simulation method and should be applicable also with other numerical approaches. While only classical models were studied here, the 2D operators ($3\times 3$
site plaquettes) used with the 3D Ising model represent a close analogy of operators typically implemented in a quantum system---operating on states
at fixed imaginary time. Here it was found that time-like correlators have an advantage over space-like ones, likely because the finite spatial extent of
the operators cause less corrections when separated in imaginary time, i.e., all the inter-plaquette operator distances are closer to the plaquette
center-center distance $r$.

An important aspect of CFTs is the fact that the conformal multiplets correspond to equally spaced scaling dimensions, with the step size equal
to $2$ when a sector of fixed Lorentz spin $l$ is considered. This fact can be very useful for testing whether there is an underlying CFT
description of a critical point observed numerically. As we have seen in the demonstrations here, the covariance method is well capable
of reproducing relevant $l=0$ primaries and their first $l=0$ descendants, and in some cases a second $l=0$ descendant can also be reasonably well
reproduced. Tests with higher $l$ incorporated in the square-shaped lattice operators (allowing for $l=1$ and $l=2$) confirm that the expected
first two or three scaling dimensions can also be reproduced there. It was also pointed out here that lattice operators that do not incorporate the
full symmetry of the lattice will in general include both $l=0$ and $l=2$ field operators, and this was tested explicitly in the 3D Ising model with
2D plaquette operators in the form of two equal dimensions from $l=0$ and $l=2$ descendants. This fact should also be kept in mind in applications
to quantum systems.

To test for a CFT corresponding to a regular critical point with a single
relevant primary, the correct spacing between their first two $l=0$ scaling dimensions already provides strong support. In light of recent progress
and remaining controversies in numerical studies pf deconfined quantum-critical points, it would be very interesting and useful to implement the
covariance method and try to extract multiple scaling dimensions in quantum MC simulations of spin and fermion models hosting such
transitions \cite{zhao20,demidio24,chen24a,chen24b,takahashi24}.

It would be interesting to apply the covariance method also to critical systems not described by CFTs, e.g., generic Mott-superfluid transitions
\cite{fisher89} or systems with quenched disorder, including classical and quantum spin glasses. The concept of conformal multiplets with their
towers of equally spaced levels does not apply there, and much less is known about the spectrum of scaling dimensions. At least on an intuitive
level, correlations with different power-law decays should correspond to different fluctuation modes and their associated scaling dimensions
should therefore be resolved in the covariance eigenvalues. Tests on various models are planned for future work.

In quantum systems, it will also be useful to study the covariance matrix as a function of imaginary time $\tau$. A method in spirit similar to
the one used here to separate different power laws was years ago employed to study exponentially decaying correlations in a system with
discrete energy levels \cite{lucher90}. This method is now often used to extract particle masses \cite{peardon97,peardon99,alpha09,liu12}. A CFT
corresponds to a gapless point, but for a finite quantum system at $T=0$, the discreteness of the level spectrum $\{\epsilon_i\}$ implies correlation
functions $C(\tau)$ mixing exponential decays $\propto {\rm e}^{-\tau\epsilon_i}$. The covariance method should resolve these levels in the same
way as in particle physics (where the method is often formulated as a generalized eigenvalue problem \cite{lucher90}), but for a CFT the
levels become more closely spaced with increasing system size, $\epsilon_i \propto \Delta_i/L$. For a sufficiently large critical system described
by a CFT, the $\tau$ dependence for $\tau$ up to order $L$ should be a sum of power laws $\tau^{-2\Delta_i}$ governed by scaling dimensions, just like
the real-space correlations analyzed here (and more generally space and time correlations are related by a dynamic exponent \cite{fisher89}). The
crossover from power-law $\tau^{-2\Delta_i}$ to exponential $\propto {\rm e}^{-a\tau\Delta_j/L}$ decays (where $a$ is a constant and $\Delta_i=\Delta_j$
does not always hold) will be studied in $d=1+1$ in a forthcoming publication \cite{lin26}. It would also be interesting to study this crossover in the
eigenvalues and -vectors of the covariance matrix in quantum MC simulations within the fuzzy sphere approach \cite{hoffmann23}, provided that
large enough system sizes can be reached to detect the power-law form (while the exponential decay should always hold asymptotically, though
with corrections to the scaling dimensions). An important question is whether the covariance method can also access CFT data beyond scaling
dimensions (central charge, coefficients of the operator-product expansion, etc.).

The closest MC based competitors to the covariance scheme would likely be MC renormalization-group (RG) methods, which have a long history
\cite{ma76,swendsen79,swendsen82}, including in the context of the same BC model studied here \cite{landau81,landau86}. This approach involves
eigenvalues of an RG transformation matrix, which also produces scaling dimensions. However, the method is less general than the covariance
matrix approach, for a given model requiring a decision on what type of RG transformation (coarse graining) to carry out. Depending on the model
this procedure may be far from obvious, especially in the case quantum systems. In contrast, the method proposed here is very general and only
requires rather trivial design of a set of operators.

Recently the MC RG approach
has been further developed by adapting neural networks (machine learning) to construct lattice realizations of continuum field operators \cite{koch18}.
Training a neural network using MC samples, some of these methods make use of covariance and principal value analysis to optimize a set of orthogonal operators 
\cite{gordon21,gokmen21a,gokmen21b,oppenheim23}. The general setup is, however, very different from, and much more complicated than, the covariance 
method demonstrated here, where the initial operators can be essentially arbitrary and their linear combinations found by diagonalization are the optimized 
operators. It would certainly be interesting to compare the performance of the methods for some challenging model.

\begin{acknowledgments}
The author would like to thank Jonathan D'Emidio, Emilie Huffman, Ami Katz, Dong-Hee Kim, Masaki Oshikawa, and Jun Takahashi for valuable
discussions and comments. This research was supported by the Simons Foundation under Grant No.~511064 and in part by Grant No.~NSF
PHY-2309135 to the Kavli Institute for Theoretical Physics (KITP). The numerical calculations were carried out on the Shared Computing
Cluster managed by Boston University's Research Computing Services.
\end{acknowledgments}

\end{document}